\documentclass[useAMS, usenatbib]{mn2e}
\usepackage{color}
\usepackage{epsfig}
\usepackage{epstopdf}
\usepackage{amsmath, amssymb}
\usepackage{hyperref}
\usepackage{wasysym} 
\usepackage{subfigure}
\usepackage{multirow}
\usepackage{lipsum}

\usepackage{setspace}  % for customisable line-spacing changes within document

\newlength{\plotwidth}
\newlength{\fullwidth}
\renewcommand{\topmargin}{-1cm} 

\usepackage{multirow}

\title[BOSS: Signs of neutrino mass in current cosmological datasets]{The clustering of galaxies in the SDSS-III Baryon Oscillation Spectroscopic Survey: Signs of neutrino mass in current cosmological datasets}
\author[Florian Beutler et al.]
{\parbox{\textwidth}{Florian Beutler$^{1}$\thanks{E-mail: \texttt{fbeutler@lbl.gov}}, Shun Saito$^{2}$, Joel R. Brownstein$^{3}$, Chia-Hsun Chuang$^{4}$, Antonio J. Cuesta$^{5}$, Will J. Percival$^{6}$, Ashley J. Ross$^{6}$, Nicholas P. Ross$^{7}$, Donald P. Schneider$^{8,9}$, Lado Samushia$^{6}$, Ariel G. S\'anchez$^{10}$, Hee-Jong Seo$^{11}$, Jeremy L. Tinker$^{12}$, Christian Wagner$^{13}$, Benjamin A. Weaver$^{12}$}\vspace{0.4cm}\\
\parbox{\textwidth}{
$^{1}$Lawrence Berkeley National Lab, 1 Cyclotron Rd, Berkeley CA 94720, USA,\\
$^{2}$Kavli Institute for the Physics and Mathematics of the Universe (WPI),Todai Institues for Advanced Study, The University of Tokyo, Chiba 277-8582, Japan,\\
$^{3}$Department of Physics and Astronomy, University of Utah, 115 S 1400 E, Salt Lake City, UT 84112, USA,\\
$^{4}$Instituto de Fisica Teorica (UAM/CSIC), Universidad Autonoma de Madrid, Cantoblanco, E-28049 Madrid, Spain,\\
$^{5}$Institut de Ciencies del Cosmos, Universitat de Barcelona, IEEC-UB, Mart\'i i Franques 1, E08028 Barcelona, Spain,\\
$^{6}$Institute of Cosmology \& Gravitation, Dennis Sciama Building, University of Portsmouth, Portsmouth, PO1 3FX, UK,\\
$^{7}$Department of Physics, Drexel University, 3141 Chestnut Street, Philadelphia, PA 19104, USA,\\
$^{8}$Department of Astronomy and Astrophysics, The Pennsylvania State University,University Park, PA 16802,\\
$^{9}$Institute for Gravitation and the Cosmos, The Pennsylvania State University,University Park, PA 16802,\\
$^{10}$Max-Planck-Institut fur extraterrestrische Physik, Postfach 1312, Giessenbachstr., 85748 Garching, Germany,\\
$^{11}$Center for Cosmology and Astroparticle Physics, Department of Physics, The Ohio State University, OH 43210, USA,\\
$^{12}$Center for Cosmology and Particle Physics, New York University, New York, NY 10003 USA,\\
$^{13}$Max-Planck-Institute for Astrophysics, Karl-Schwarzschild-Str. 1, 85748 Garching, Germany.}}

\begin{document}

\label{firstpage}

\maketitle

\begin{abstract}
We investigate the cosmological implications of the latest growth of structure measurement from the Baryon Oscillation Spectroscopic Survey (BOSS) CMASS Data Release 11 with particular focus on the sum of the neutrino masses, $\sum m_{\nu}$. We examine the robustness of the cosmological constraints from the Baryon Acoustic Oscillation (BAO) scale, the Alcock-Paczynski effect and redshift-space distortions ($D_V/r_s$, $F_{\rm AP}$, $f\sigma_8$) of~\citet{Beutler:2013b}, when introducing a neutrino mass in the power spectrum template.
We then discuss how the neutrino mass relaxes discrepancies between the Cosmic Microwave Background (CMB) and other low-redshift measurements within $\Lambda$CDM.
Combining our cosmological constraints with WMAP9 yields $\sum m_{\nu} = 0.36\pm0.14\,$eV ($68\%$ c.l.), which represents a $2.6\sigma$ preference for non-zero neutrino mass. The significance can be increased to $3.3\sigma$ when including weak lensing results and other BAO constraints, yielding $\sum m_{\nu} = 0.35\pm0.10\,$eV ($68\%$ c.l.). However, combining CMASS with Planck data reduces the preference for neutrino mass to $\sim 2\sigma$.
When removing the CMB lensing effect in the Planck temperature power spectrum (by marginalising over $A_{\rm L}$), we see shifts of $\sim 1\sigma$ in $\sigma_8$ and $\Omega_m$, which have a significant effect on the neutrino mass constraints. In case of CMASS plus Planck without the $A_{\rm L}$-lensing signal, we find a preference for a neutrino mass of $\sum m_{\nu} = 0.34\pm 0.14\,$eV ($68\%$ c.l.), in excellent agreement with the WMAP9+CMASS value. The constraint can be tightened to $3.4\sigma$ yielding $\sum m_{\nu} = 0.36\pm 0.10\,$eV ($68\%$ c.l.) when weak lensing data and other BAO constraints are included. \vspace{1cm}
\end{abstract}


\section{introduction}

The measurement of neutrino oscillations in neutrino detection experiments using solar, atmospheric and reactor neutrinos has now convincingly shown that neutrinos cannot be massless. 
Neutrino oscillation experiments are sensitive to the mass differences between the neutrino eigenstates, and the current data imply $|\Delta m^2_{31}| \cong 2.4\times 10^{-3}$eV$^2$ and $\Delta m^2_{21} \cong 7.6\times 10^{-5}$eV$^2$~\citep{Beringer:1900zz}. These measurements provide a lower limit for the sum of the neutrino masses of $\sim 0.06\,$eV. Using the mass difference constraints above and knowing that $\Delta m^2_{21} > 0$, one can construct two mass hierarchies for neutrinos. The so-called ``normal'' hierarchy suggests $m_{\nu_1} < m_{\nu_2} \ll m_{\nu_3}$, where we have one heavy neutrino and two lighter ones, while the so-called ``inverted'' hierarchy suggests $m_{\nu_3} \ll m_{\nu_1} < m_{\nu_2}$, where we have one light neutrino and two heavy ones. 

Because of the extremely low cross-section of neutrinos it is difficult for laboratory experiments to measure the neutrino mass directly.
The current best upper bounds on the neutrino mass from particle physics experiments are from Troitsk~\citep{Lobashev:1999tp} and Mainz~\citep{Weinheimer:1999tn} tritium beta-decay experiments that found $m_{\beta} < 2.3\,$eV ($95\%$ confidence level), where $m_{\beta}$ is the mass to which beta-decay experiments are sensitive (see section~\ref{sec:particle} and eq.~\ref{eq:mbeta}). The KArlsruhe TRItium Neutrino experiment (KATRIN) aims to measure $m_{\beta}$ with a sensitivity of $\sim 0.2\,$eV~\citep{Wolf:2008hf}, which would constrain $\sum m_{\nu} \lesssim 0.6\,$eV. 
Also, neutrino-less double beta decay ($0\nu\beta\beta$) experiments such as KamLAND-Zen will assess the effective mass of Majorana neutrinos at the level of $\mathcal{O}$($0.1-1$)eV~\citep{Gando:2012zm} depending on the nuclear matrix element.

With the advent of precision cosmology, it was realised that the neutrino mass has an effect on the matter distribution in the Universe and that this could be used to indirectly measure the sum of the neutrino masses, $\sum m_{\nu}$. The neutrino mass introduces a scale-dependent suppression of the clustering amplitude with the scale-dependency set by $f_{\nu} = \Omega_{\nu}/\Omega_m$. The suppression of clustering is caused by the large thermal velocity of neutrinos which leads to a large free-streaming scale. Many recent publications have attempted to constrain $\sum m_{\nu}$, but most were only able to set upper limits~\citep{Seljak:2006bg, Hinshaw:2008kr,Dunkley:2008ie,Reid:2009nq,Komatsu:2010fb,Saito:2010pw,Thomas:2009ae,Tereno:2008mm,Gong:2008pg,Ichiki:2008ye,Li:2008vf,Zhao:2012xw,Hinshaw:2012aka,dePutter:2012sh,Xia:2012na,Sanchez:2012sg,Riemer-Sorensen:2013jsa,Giusarma:2013pmn} with some exceptions based on cluster abundance results, e.g.,~\citet{Hou:2012xq,Ade:2013lmv,Battye:2013xqa,Wyman:2013lza,Burenin:2013wg,Rozo:2013hha}. 

Introducing a neutrino mass suppresses clustering power between the epoch of decoupling and today below the free streaming scale, as massive neutrinos affect the cosmological expansion rate, but free-stream out of matter perturbations. The clustering amplitude is often parameterised by the r.m.s. mass fluctuations in spheres of $8\,$Mpc$/h$ at the present epoch and denoted $\sigma_8$. Given the clustering amplitude at decoupling measured by the CMB, we can predict the $z=0$ value of $\sigma_8$, within a certain cosmological model. However, this $\sigma_8$ prediction depends on the initial assumption of the neutrino mass, introducing a degeneracy between $\sigma_8$ and $\sum m_{\nu}$.
In fact, if there were no other effect of the neutrino mass on the CMB, the neutrino mass parameter would be completely degenerate with $\sigma_8$. Luckily there are several other effects of the neutrino mass on the CMB, which can be used to break this degeneracy. If neutrinos would exceed the limit $\sum m_{\nu} \lesssim 1.8\,$eV, they would trigger more direct effects in the CMB~\citep{Dodelson:1995es,Ichikawa:2004zi}, which are not observed. This represents probably the most robust limit on the neutrino mass from cosmology. Apart from this there are other, more subtle effects on the CMB anisotropies.
Changing the neutrino mass and keeping the redshift of matter-radiation equality fixed will change the low-redshift value of $\Omega_mh^2$. This will change the angular diameter distance to the last scattering surface, $D_A(z_*)$. Since such changes can be absorbed by changes in the Hubble parameter there is a (geometric) degeneracy between $\sum m_{\nu}$ and $H_0$ in the CMB. Beside the angular diameter distance the neutrino mass also impacts the slope of the CMB power spectrum at low multipoles due to the Integrated Sachs-Wolfe (ISW) effect~\citep{Lesgourgues:2006nd,Ade:2013dsi}. The ISW effect describes the energy change of CMB photons caused by the decay of the gravitational potentials during radiation domination (early ISW effect) or $\Lambda$ domination (late ISW effect). If instead $\Omega_mh^2$ is kept fixed when varying the neutrino mass, the redshift of matter-radiation equality will change, which affects the position and amplitude of the acoustic peaks in the CMB power spectrum (for more details see e.g.~\citealt{oai:arXiv.org:1212.6154}). 
Weak gravitational lensing of the CMB photons encodes information about the late-time Universe with the Planck kernel peaking at around $z\sim 2$~\citep{Ade:2013aro}. The lensing deflections are caused by an integrated measure of the matter distribution along the line of sight. 
Using these additional signals, the CMB data are able to break the $\sum m_{\nu}$-$\sigma_8$ degeneracy to some extent. The remaining degeneracy can be broken by including low-redshift $\sigma_8$ measurements from other datasets.

Low-redshift measurements of the clustering amplitude ($\sigma_8$) are notoriously difficult, and to some extent require priors from the CMB. Most low-redshift probes which are sensitive to $\sigma_8$ require the understanding of non-linear effects and usually carry large systematic uncertainties. In this paper we demonstrate that recent constraints on the growth rate $f\sigma_8$, the Baryon Acoustic Oscillation (BAO) scale $D_V/r_s$ and the Alcock-Paczynski effect $F_{\rm AP}$ from the Baryon Oscillation Spectroscopic Survey (BOSS) are robust against variations of $\sum m_{\nu}$ in the theoretical template. We also show that the constraint on ($\Omega_m$, $\sigma_8$) from the shear correlation function of the weak lensing signal of CFHTLenS is robust against variations of $\sum m_{\nu}$. We therefore claim that combining CMB datasets with these low-redshift growth of structure measurements provides a reliable approach to break the $\sum m_{\nu}$-$\sigma_8$ degeneracy in the CMB.

This paper is organised as follows. In section~\ref{sec:theory} we give a brief summary of the effect of neutrinos on the matter perturbations. In section~\ref{sec:tension} we introduce constraints on $\sigma_8$ from different datasets. In section~\ref{sec:reliability} we investigate the robustness of different low redshift $\sigma_8$ constraints, with particular focus on the BOSS growth of structure constraint. In section~\ref{sec:mnu} we investigate the parameter $\sum m_{\nu}$ as one approach to relieve the tension between these different datasets. In section~\ref{sec:dis} we discuss the cosmological implications of our results, and we conclude in section~\ref{sec:conclusion}.

\section{Background}
\label{sec:theory}

Here we give a brief overview of the effect of massive neutrinos on the matter perturbations in the Universe. More details can be found in most standard text books (see also~\citealt{Lesgourgues:2006nd,oai:arXiv.org:1212.6154,Lbook}). 

In the absence of massive neutrinos, density perturbations of cold dark matter and baryons grow as
\begin{equation}
\delta_{\rm m}\propto \begin{cases} a & \text{during matter domination},\\
aD(a) & \text{during }\Lambda\text{ domination},\end{cases}
\end{equation} 
where $a$ is the scale factor and $D(a)$ is the growth function. 
The large thermal velocity of neutrinos leads to a free streaming scale, below which they do not contribute to the matter perturbation growth. During matter or $\Lambda$ domination the free streaming scale can be approximated by
\begin{equation}
k_{\rm FS} = 0.82\frac{\sqrt{\Omega_{\Lambda} + \Omega_m(1+z)^3}}{(1+z)^2}\left(\frac{m_{\nu}}{1\,\text{eV}}\right)h/\text{Mpc},
\label{eq:kfs}
\end{equation}
where $m_{\nu}$ is the mass of the individual neutrino eigenstates. 
The matter growth is an interplay between the dilution of matter caused by the expansion of the universe and the growth of perturbations through gravitational collapse. Since neutrinos contribute to the homogeneous expansion through the Friedmann equation but do not contribute to the growth of matter perturbations below the free streaming scale, the overall growth of cold dark matter and baryon perturbations on small scales is suppressed and behaves approximately as 
\begin{equation}
\delta_{\rm cb}\propto \begin{cases} a^{1-\frac{3}{5}f_{\nu}} & \text{during matter domination},\\
\left[aD(a)\right]^{1-\frac{3}{5}f_{\nu}} & \text{during }\Lambda\text{ domination}\end{cases}
\end{equation}
with $f_{\nu} = \Omega_{\nu}/\Omega_m$~\citep{Bond:1980ha}. The total matter perturbations are then given by $\delta_m = (1-f_{\nu})\delta_{cb} + f_{\nu}\delta_{\nu}$. 
The suppression of matter perturbations on small scales leads to a suppression of the power spectrum, $P = \langle \delta\delta^*\rangle$. In the linear regime this suppression can be approximated by~\citep{Hu:1997mj}
\begin{equation}
\frac{P^{f_{\nu}}-P^{f_{\nu}=0}}{P^{f_{\nu}=0}} \simeq -8f_{\nu},
\end{equation}
while on scales larger than the free streaming scale, neutrino perturbations behave like cold dark matter (see also~\citealt{Brandbyge:2008rv,Viel:2010bn}).

The overall normalisation of the power spectrum is usually parameterised by $\sigma_8$ with $P \propto \sigma_8^2$. The CMB measures the scalar amplitude $A_s$, which must be extrapolated from the redshift of decoupling $z_* \approx 1100$ to redshift zero to obtain $\sigma_8$. The relation between $A_s$ and $\sigma_8$ is given by
\begin{align}
\sigma_8^2(a) &\propto A_s \int^{\infty}_{0}dk\,k^2D^2(a)T^2(k)k^{n_s}W(kR),
\label{eq:sig8z}
\end{align}
where $R = 8h^{-1}\,$Mpc, $n_s$ is the scalar spectral index and $W(x) = 3[\sin(x) - x\cos(x)]/x$ is the Fourier transform of the top-hat window function. The growth factor $D(a)$, the primordial power spectrum $k^{n_s}$ and the transfer function $T(k)$ define the power spectrum up to a normalisation factor, $P(k,a) \propto D^2(a)k^{n_s}T^2(k)$.
Comparing low redshift $\sigma_8$ measurements with the extrapolation of $\sigma_8$ from the CMB allows us to measure the damping effect of neutrinos~\citep{Takada:2005si}. 
In this paper we are going to use constraints from galaxy redshift surveys as well as weak lensing. While galaxy surveys measure the galaxy power spectrum, which can be related to the matter power spectrum by the galaxy bias, weak lensing surveys are sensitive to a line-of-sight integral over the matter power spectrum weighted by a lensing kernel.

\section{Cosmological datasets included in this analysis}
\label{sec:tension}

Here we introduce the different datasets used in our analysis. We start with the Baryon Oscillation Spectroscopic Survey (BOSS) CMASS sample. BOSS, as part of SDSS-III~\citep{Eisenstein:2011sa,Dawson:2012va} is measuring spectroscopic redshifts of $\approx 1.5$ million galaxies (and $150\,000$ quasars) using the SDSS multi-fibre spectrographs~\citep{Bolton:2012hz,Smee:2012wd}. The galaxies are selected from multi-colour SDSS imaging~\citep{Fukugita:1996qt,Gunn:1998vh,Smith:2002pca,Gunn:2006tw,Doi:2010rf} and cover a redshift range of $z = 0.15$ - $0.7$, where the survey is split into two samples called LOWZ ($z=0.15$ - $0.43$) and CMASS ($z=0.43$ - $0.7$). The CMASS-DR11 sample covers $6\,391\deg^2$ in the North Galactic Cap and $2\,107\deg^2$ in the South Galactic Cap; the total area of $8\,498\deg^2$ represents a significant increase from CMASS-DR9~\citep{Ahn:2012fh,Anderson:2013oza}, which covered $3\,265\deg^2$ in total. In this analysis we use the CMASS-DR11 results of~\citet{Beutler:2013b}, which includes the measurement of the BAO scale, the Alcock-Paczynski effect and the signal of redshift-space distortions (RSD). Note that we do not include other RSD measurements (e.g.~\citealt{Hawkins:2002sg,Blake:2011rj,Beutler:2012px,Oka:2013cba}), since the measurement methodology and nonlinear modelling are different from the one in~\citet{Beutler:2013b}.

We also use results from SDSS-II (DR7)~\citep{Abazajian:2008wr} reported in~\citet{Mandelbaum:2012ay}, where galaxy-galaxy weak lensing together with galaxy clustering has been used to constrain the dark matter clustering.

Our second lensing dataset is the Canada-France-Hawaii Lensing Survey~\citep{Heymans:2012gg}, hereafter referred to as CFHTLenS. The CFHTLenS survey analysis combined weak lensing data processing with {\sc theli}~\citep{Erben:2012zw}, shear measurement with {\sc lensfit}~\citep{Miller:2012am}, and photometric redshift measurement with PSF-matched photometry~\citep{Hildebrandt:2011hb}. A full systematic error analysis of the shear measurements in combination with the photometric redshifts is presented in~\citet{Heymans:2012gg}, with additional error analyses of the photometric redshift measurements presented in~\citet{Benjamin:2012qp}.

The current most powerful cosmological datasets are measurements of the Cosmic Microwave Background (CMB). We include the $9$-year results from the Wilkinson Microwave Anisotropy Probe (WMAP)~\citep{Hinshaw:2012aka} and the first data release of Planck~\citep{Ade:2013zuv} interchangeably. Both datasets have full-sky, multi-frequency coverage and the Planck beams are small enough to probe gravitational lensing deflections which was not possible in WMAP. We will also compare the original results of Planck with the re-analysis by~\citet{Spergel:2013rxa}.

\begin{figure}
\begin{center}
\epsfig{file=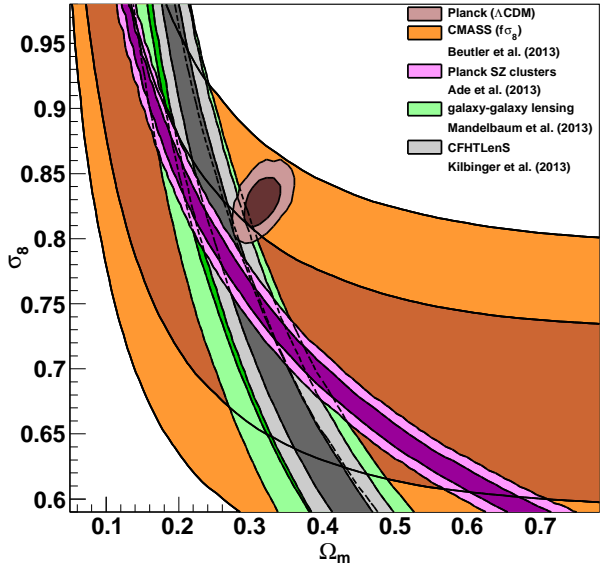,width=8.8cm}
\caption{Comparison between the likelihood distributions in $\Omega_m$-$\sigma_8$ within $\Lambda$CDM. We show Planck~\citep{Ade:2013zuv} (brown contours), Planck SZ clusters~\citep{Ade:2013lmv} (magenta contours), CFHTLenS~\citep{Kilbinger:2012qz} (grey contours), galaxy-galaxy lensing~\citep{Mandelbaum:2012ay} (green contours) and CMASS-DR11~\citep{Beutler:2013b} (orange contours). The Planck contours in this plot assume $\Lambda$CDM and $\sum m_{\nu} = 0.06\,$eV.}
\label{fig:om_sig8}
\end{center}
\end{figure}

In Figure~\ref{fig:om_sig8} we present two-dimensional likelihood distributions of $\Omega_m$ and $\sigma_8$ from the datasets discussed above. Within a flat $\Lambda$CDM cosmological model the CMB provides by far the best constraints in this parameter space (brown contours). We show the Planck+WP result, which includes the polarisation analysis from WMAP. 
The Planck prediction of $\Omega_m$ and $\sigma_8$ relies strongly on the assumption of $\Lambda$CDM, since both parameters, $\Omega_m$ and $\sigma_8$, are extrapolated from information at high redshift. We compare the Planck prediction with the lensing result from the CFHTLenS~\citep{Kilbinger:2012qz} (grey contours), the galaxy-galaxy lensing result of~\citet{Mandelbaum:2012ay} (green contours) and the result using clusters detected through the Sunyaev-Zel'dovich (SZ) effect in Planck~\citep{Ade:2013lmv} (magenta contours). The orange contours show the CMASS-DR11 results of~\citet{Beutler:2013b} in the form of $f\sigma_8$ (see also ~\citealt{Samushia:2013yga,Sanchez:2013tga,Chuang:2013wga}, which gave very similar results). The largest disagreement with the Planck prediction comes from the SZ cluster result and has been discussed in~\citet{Ade:2013lmv} (see also~\citealt{Benson:2011uta} for similar results from SPT). 

\section{The reliability of low redshift growth of structure constraints}
\label{sec:reliability}

We now investigate the reliability of different low redshift growth of structure measurements with respect to variations of $\sum m_{\nu}$ in the data modelling process. Low redshift growth of structure measurements usually report some combined constraint on $\Omega_m$ and $\sigma_8$ and assume a certain value for the neutrino mass when deriving this constraint. To be able to use such a measurement to set limits on the value of the neutrino mass one needs to be sure that the initial assumption about the neutrino mass in the modelling step does not influence the result. 

The tightest current constraint comes from SZ clusters detected by Planck~\citep{Ade:2013lmv} in the form:
\begin{equation}
\sigma_8\left(\frac{\Omega_m}{0.27}\right)^{0.3} = 0.78\pm 0.01.
\end{equation}
We include this result as the magenta contours in Figure~\ref{fig:om_sig8}. The tension between this measurement and the Planck temperature power spectrum has been discussed in~\citet{Ade:2013lmv} (see also~\citealt{Hamann:2013iba,Battye:2013xqa}); and including a large value for the sum of the neutrino masses is mentioned as one possibility to relieve this tension. However, \citet{Costanzi:2013bha} showed that in the case of $\sum m_{\nu} = 0.4\,$eV there is up to $30\%$ uncertainty on the predicted cluster count depending on whether the r.m.s. of the mass perturbations, $\sigma(M)$, required to predict the halo mass function, is calculated from the cold dark matter power spectrum or the matter power spectrum. In case of the Planck SZ cluster analysis the systematic uncertainty on $\sigma_8$ from these effects is twice as large as the statistical error reported by the Planck collaboration. Cluster counts also carry an uncertainty from the unknown bias in the mass-observable relations~\citep{Rozo:2013hha,vonderLinden:2014haa}. Because of these uncertainties we will not include the SZ cluster results in the parameter constraints in this paper.

\subsection{Reliability of the CMASS constraints}

\begin{figure}
\begin{center}
\epsfig{file=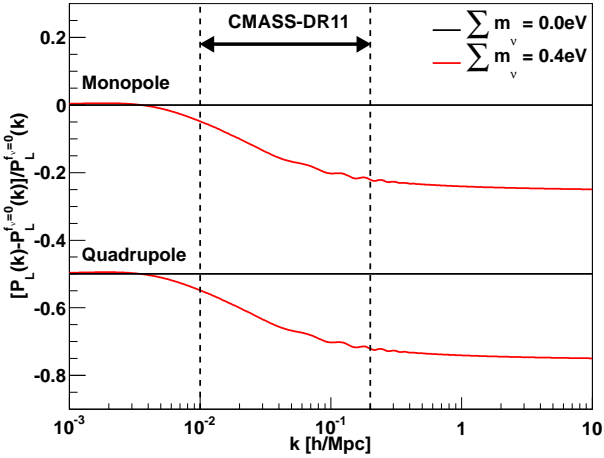,width=8.8cm}
\caption{Relative amplitude difference between a linear power spectrum monopole (top) and quadrupole (bottom) with $\sum m_{\nu} = 0\,$eV (black lines) and $\sum m_{\nu} = 0.4\,$eV (red lines). We keep $\Omega_ch^2$ fixed when including the neutrino mass, so that the total physical matter density increases as $\Omega_mh^2 = \Omega_ch^2 + \Omega_bh^2 + \Omega_{\nu}h^2$. The black dashed lines show the fitting range for the CMASS-DR11 results of~\citet{Beutler:2013b}. We subtract $0.5$ from the quadrupole for plotting purposes.}
\label{fig:psmnu}
\end{center}
\end{figure}

\begin{table*}
\begin{center}
\caption{Comparison between the best fitting and mean parameters of~\citet{Beutler:2013b} and the results obtained in this paper, where $\sum m_{\nu} = 0.4\,$eV has been used in the power spectrum template. The first three rows show the main cosmological parameters, while the last four rows show the $4$ nuisance parameters of the fit. The fitting range is $k=0.01$ - $0.20h/$Mpc. Details about the modelling can be found in~\citet{Beutler:2013b}.}
	\begin{tabular}{cllll}
		\hline
		& \multicolumn{2}{c}{Beutler et al. 2013} & \multicolumn{2}{c}{Template with $\sum m_{\nu} = 0.4\,$eV}\\
		parameter & best fit & mean & best fit & mean\\
		\hline
		$D_V(z_{\rm eff})/r_s(z_d)$ & $13.88$ & $13.89\pm 0.18$ & $13.87$ & $13.91\pm 0.25$\\
		$F_{\rm AP}(z_{\rm eff})$ & $0.683$ & $0.679\pm 0.031$ & $0.664$ & $0.664\pm 0.035$\\
		$f(z_{\rm eff})\sigma_8(z_{\rm eff})$ & $0.422$ & $0.419\pm 0.044$ & $0.404$ & $0.405\pm 0.048$\\
		\hline
		$b_1\sigma_8(z_{\rm eff})$ & $1.221$ & $1.224\pm 0.031$ & $1.183$ & $1.188\pm 0.039$\\
		$b_2\sigma_8(z_{\rm eff})$ & $-0.21$ & $-0.09\pm 0.62$ & $-0.67$ & $-0.72\pm 0.41$\\
		$\sigma_v$ & $4.63\,$Mpc$/h$ & $4.65\pm 0.81\,$Mpc$/h$ & $4.9\,$Mpc$/h$ & $4.9\pm 1.0\,$Mpc$/h$\\
		$N$ & $1890\,[\text{Mpc}/h]^3$ & $1690\pm 600\,[\text{Mpc}/h]^3$ & $3400\,[\text{Mpc}/h]^3$ & $3400\pm 1100\,[\text{Mpc}/h]^3$\\
		\hline
	  \end{tabular}
	  \label{tab:cmp}
\end{center}
\end{table*}

\begin{figure}
\begin{center}
\epsfig{file=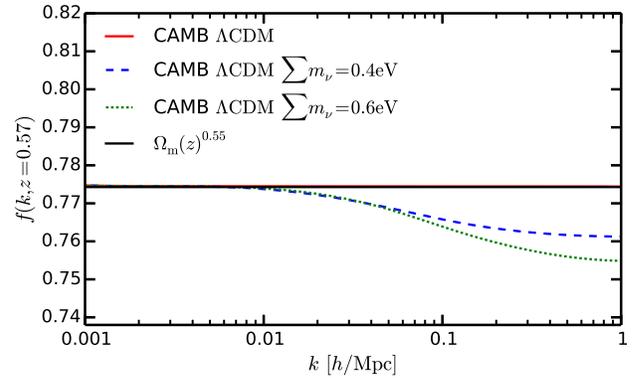,width=8.8cm}
\caption{Scale-dependence of the growth rate for different values of the neutrino mass parameter. The black line shows the commonly used linear assumption, while all other lines are obtained as derivatives of the growth factor $D(k, a)$ using a {\sc camb} power spectrum. In this figure we fix $\Omega_mh^2$ when increasing the neutrino mass.}
\label{fig:dlnDdlna_comp}
\end{center}
\end{figure}

In \citet{Beutler:2013b} we analysed the CMASS power spectrum multipoles and constrained the distance ratio $D_V/r_s(z_d)$, where $r_s(z_d)$ is the sound horizon at the drag epoch, and 
\begin{equation}
D_V = \left[(1+z)^2D^2_A(z)\frac{cz}{H(z)}\right]^{1/3},
\end{equation}
as well as the Alcock-Paczynski parameter $F_{\rm AP} = (1+z)D_A(z)H(z)/c$ and the growth of structure $f\sigma_8$. Our technique made use of a power spectrum template, based on the current Planck cosmological parameters including several non-linear effects which are parameterised by four nuisance parameters. We scaled this template with two scaling parameters, $\alpha_{\parallel}$ and $\alpha_{\perp}$, along with the redshift-space distortion parameter $f\sigma_8$. We performed several systematics tests, which demonstrated that the power spectrum can be used up to $k=0.20h/$Mpc, where our constraints are still dominated by the statistical error. 
To some extent our technique must be understood as a consistency check within the Planck cosmological parameters, since we rely on the Planck power spectrum template. However, as shown in~\citet{Beutler:2013b}, the parameter constraints do not rely heavily on the template itself. For example, we can replace the Planck template with a WMAP9 template and not bias our parameter constraints (see~\citealt{Beutler:2013b} section $7$ for details). Another systematics check has been performed in~\citet{Ross:2013vla}, where we showed that the BAO and redshift-space distortion constraints are independent of different colour cuts within the CMASS dataset.

\begin{figure*}
\begin{center}
\epsfig{file=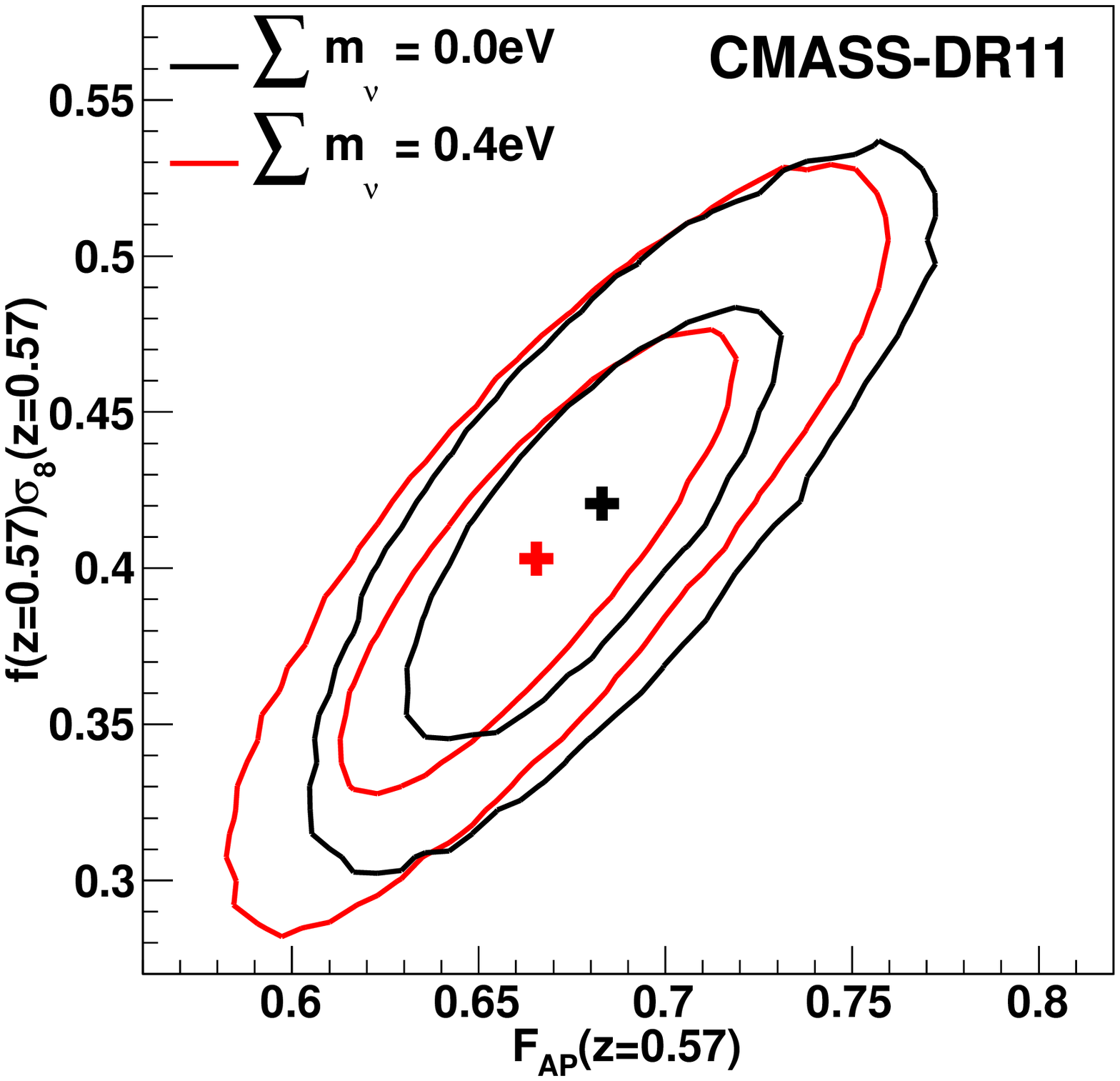,width=8.8cm}
\epsfig{file=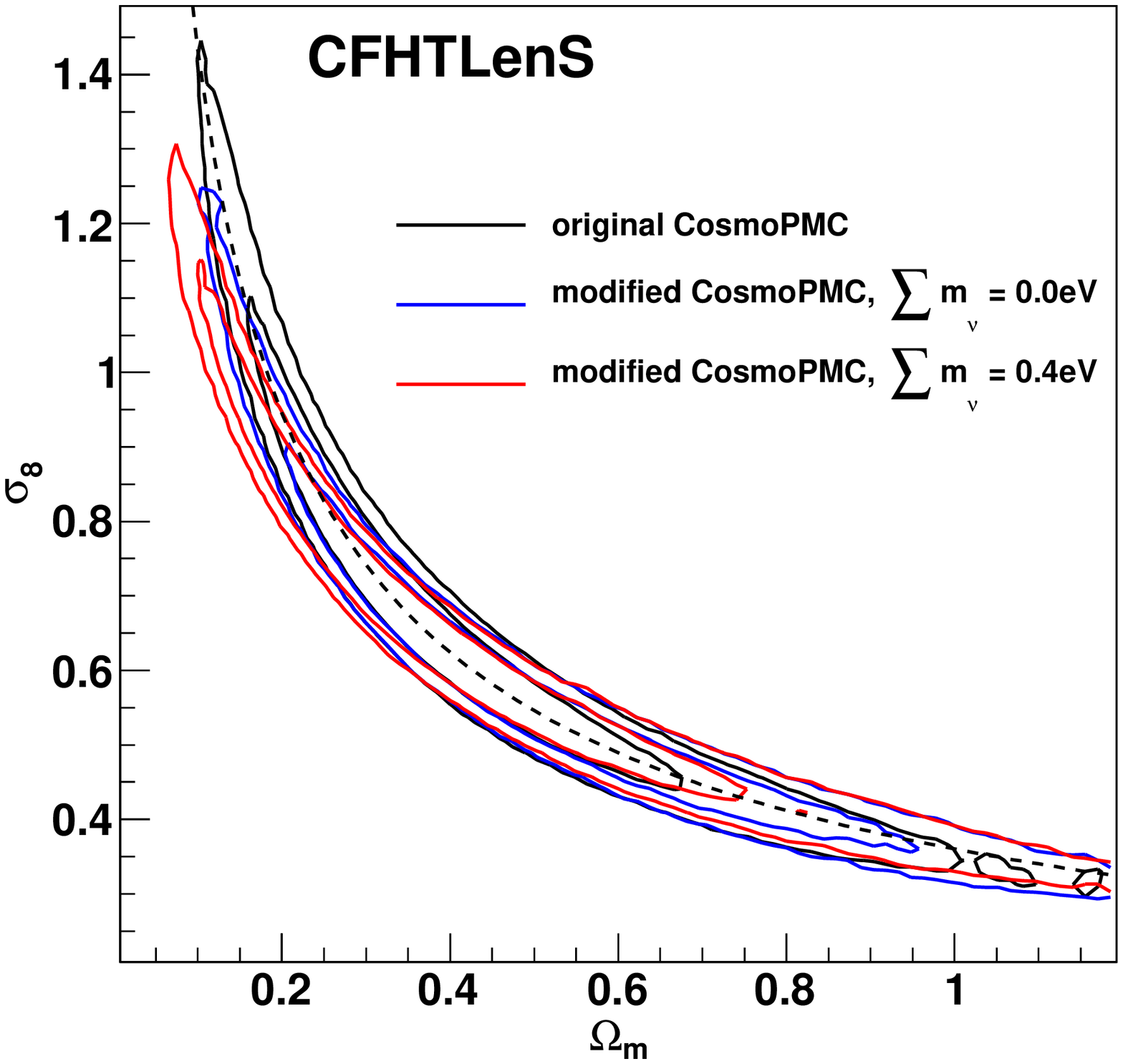,width=8.8cm}
\caption{Results of the reliability tests for the CMASS and CFHTLenS constraints. The red contours include a neutrino mass of $\sum m_{\nu} = 0.4\,$eV in the modelling, while the black contours assume $\sum m_{\nu} = 0\,$eV. (left) Here we show the Alcock-Paczynski parameter $F_{\rm AP}$ and the growth rate $f\sigma_8$ from CMASS-DR11, which are the two parameters most affected by the change in the neutrino mass parameter (in the analysis we also include the BAO scale via $D_V/r_s$). The crosses mark the maximum likelihood values. (right) Here we show the $\Omega_m$-$\sigma_8$ constraints of CFHTLenS including the degeneracy line used in our analysis and reported in~\citet{Kilbinger:2012qz} (black dashed line). The black contours show the original fitting results using the CosmoPMC implementation of~\citep{Kilbinger:2011bu}, while the blue contours use a modified code with the biggest difference being the new {\sc halofit} implementation of~\citet{Bird:2011rb} (see text for details).}
\label{fig:reltests}
\end{center}
\end{figure*}

The question we want to address here is, what happens when we introduce a neutrino mass in the power spectrum template? The power spectrum template in~\citet{Beutler:2013b} assumes $\sum m_{\nu} = 0\,$eV. The neutrino mass introduces a scale-dependent damping in the power spectrum. Since our analysis includes four nuisance parameters to capture scale-dependencies, we can expect that some of the changes in the power spectrum template will be absorbed by these nuisance parameters. 

To explicitly test the effect of the neutrino mass we produce a linear power spectrum using {\sc camb}~\citep{Lewis:2002ah} setting the neutrino mass parameter to $\sum m_{\nu} = 0.4\,$eV with three massive neutrinos with degenerate masses. 
We keep $\Omega_ch^2$ fixed when including the neutrino mass, so that the total physical matter density increases as $\Omega_mh^2 = \Omega_ch^2 + \Omega_bh^2 + \Omega_{\nu}h^2$. The physical neutrino density is given by 
\begin{equation}
\Omega_{\nu}h^2 = \frac{\sum m_{\nu}}{93.14\,\text{eV}}.
\end{equation}
We use the linear matter power spectrum as input for the non-linear power spectrum produced via {\sc RegPT}~\citep{Taruya:2012ut} as well as the correction terms summarised in section $6$ of~\citet{Beutler:2013b}. Strictly speaking, {\sc RegPT} is not designed to include the non-linear clustering contribution of neutrinos. By using the linear matter power spectrum as an input to {\sc RegPT} we assume that non-linear corrections for the matter power spectrum are identical for neutrinos, cold dark matter and baryons. This assumption is incorrect, since neutrino perturbations tend to stay in the linear regime even where non-linear corrections to the matter power spectrum are not negligible~\citep{Saito:2008bp,Wong:2008ws,Saito:2009ah}. However, these effects are small on the scales we are interested in and should not influence the outcome of this test. The difference between the linear matter power spectrum multipoles with a neutrino mass parameter of $\sum m_{\nu} = 0.4\,$eV and zero neutrino mass is shown in Figure~\ref{fig:psmnu}.

Also notice that a non-zero neutrino mass introduces a scale-dependence in the growth rate $f(k, a) = d\ln D(k,a)/d\ln a$ as shown in Figure~\ref{fig:dlnDdlna_comp}. To compare with the CMB prediction of the growth rate we desire to know the value $f(k\rightarrow 0)$ instead of some effective growth rate. In case of $\sum m_{\nu} = 0.4\,$eV the suppression is $1.4\%$ at $k=0.20h/$Mpc. We include this effect in our parameter fit, by using $f\sigma_8g(k)$ as a free parameter instead of $f\sigma_8$, where
\begin{equation}
g(k) = \frac{f(z,\sum m_{\nu} = 0.4\,\text{eV})}{f(z,\sum m_{\nu} = 0\,\text{eV})}.
\end{equation}
Using the new power spectrum template we repeat the parameter fit of~\citet{Beutler:2013b} with the fitting range $k = 0.01$ - $0.20h/$Mpc. Table~\ref{tab:cmp} and Figure~\ref{fig:reltests} (left) summarise the results. The new template is a slightly worse fit to the dataset compared to the template with $\sum m_{\nu} = 0\,$eV used in~\citet{Beutler:2013b}, since the $\chi^2$ increases by $\Delta \chi^2 = 7.6$ to $\chi^2/$d.o.f. $= 148.1/145$. This degradation is also reflected in the increased errorbars for some parameters. However, this result cannot be interpreted as preference for zero neutrino mass, since all other cosmological parameters in the power spectrum template were fixed. The best fitting values show shifts $< 0.5\sigma$ for the three cosmological parameters. As expected, the BAO scale is very robust against changes in the template, while we see larger shifts in $F_{\rm AP}$ and $f\sigma_8$ (see the Figure~\ref{fig:reltests}, left). Since $F_{\rm AP}$ and $f\sigma_8$ are correlated the significance of this shift is lower when accounting for the correlation. We calculate the quantity
\begin{equation}
\Delta \chi^2_{\rm dif} = (V^{\rm data}_{f_{\nu}} - V^{\rm data}_{f_{\nu} = 0})^TC^{-1}(V^{\rm data}_{f_{\nu}} - V^{\rm data}_{f_{\nu} = 0}),
\end{equation}
where $V^{\rm data}_{f_{\nu}}$ is the data vector of $D_V/r_s$, $F_{\rm AP}$ and $f\sigma_8$, derived in this paper and $V^{\rm data}_{f_{\nu} = 0}$ stands for the equivalent results of~\citet{Beutler:2013b}. We use the covariance matrix, $C^{-1}$, reported in eq.~$72$ of~\citet{Beutler:2013b}. For the values in Table~\ref{tab:cmp} we find $\Delta \chi^2_{\rm dif} = 0.29$ with three degrees of freedom. While the shifts reported in Table~\ref{tab:cmp} do not seem significant, they lead to smaller $f\sigma_8$ which would increase the preference for a neutrino mass, as discussed in the next section. These possible systematics are of the same order as the modelling systematics discussed in~\citet{Beutler:2013b}, which when treated as independent error contributions increase the total error budged by $5\%$. We conclude that our cosmological parameter constraints are robust against changes in the power spectrum template related to the neutrino mass. 

\subsection{Reliability of the CFHTLenS constraint}

Now we test the reliability of the CFHTLenS results when changing the neutrino mass parameter in the shear-shear correlation function model. We use the Population Monte Carlo code {\sc CosmoPMC}~\citep{Kilbinger:2011bu}, which was also used for the CFHTLenS analysis in~\citet{Kilbinger:2012qz}. {\sc CosmoPMC} uses the linear matter power spectrum fitting formula of~\citet{Eisenstein:1997ik} when modelling the shear-shear correlation function as well as the {\sc halofit} mapping of~\citet{Smith:2002dz}, neither of which includes the effect of massive neutrinos. We therefore modify the code by including the linear power spectrum fitting formula of~\citet{Eisenstein:1997jh} and the {\sc halofit} mapping suggested in~\citet{Bird:2011rb}. The {\sc halofit} implementation of~\citet{Bird:2011rb} also includes a correction to the power spectrum amplitude on small scales, which~\citet{Smith:2002dz} over-predicts. The effect of this correction to the CFHTLenS constraints is shown in Figure~\ref{fig:reltests} (right), where the black contours use the {\sc halofit} mapping of~\citet{Smith:2002dz}, while the blue contours use the {\sc halofit} mapping of~\citet{Bird:2011rb}. If we also include a neutrino mass of $\sum m_{\nu} = 0.4\,$eV we obtain the red contours. At $\Omega_m = 0.3$ the difference between black and the blue contours is $\sim 1\sigma$ for $\sigma_8$. When introducing a neutrino mass (red contours) there is another shift of $\sim \sigma/3$. We therefore conclude that while the CFHTLenS constraint shows some sensitivity to the exact treatment of non-linear clustering, it does not seem to be very sensitive to the effect of the neutrino mass. Note that all shifts shown in Figure~\ref{fig:reltests} (right) lead to a smaller clustering amplitude (at fixed $\Omega_m$) and therefore would increase the preference for neutrino mass (see next section).

Here we are not explicitly testing the reliability of the galaxy-galaxy lensing result of~\citet{Mandelbaum:2012ay} which we are also going to use later in this analysis, but refer the reader to section 2.3.2 in~\citet{Mandelbaum:2012ay}.

\section{Constraining the mass of neutrinos}
\label{sec:mnu}

\begin{table*}
\begin{center}
\caption{Constraints on $\sigma_8$, $\Omega_m$ and $\sum m_{\nu}$ combining different datasets. The errors on $\sigma_8$ and $\Omega_m$ are $1\sigma$, while for $\sum m_{\nu}$ we report the $68\%$ and $95\%$ confidence levels. Planck stands for the Planck+WP result reported in~\citet{Ade:2013zuv}, WMAP9 represents the 9-year results of WMAP reported in~\citet{Hinshaw:2012aka}, Spergel2013 stands for the Planck re-analysis of~\citet{Spergel:2013rxa}, Beutler2013 represents the constraints on $D_V/r_s$, $F_{\rm AP}$ and $f\sigma_8$ from~\citet{Beutler:2013b}, CFHTLenS represents the weak lensing results from~\citet{Kilbinger:2012qz}, GGlensing represents the galaxy-galaxy lensing results reported in~\citet{Mandelbaum:2012ay} and BAO stands for the BAO constraint of 6dFGS~\citep{Beutler:2011hx} and the isotropic BAO constraints of LOWZ~\citep{Anderson2.0,Tojeiro:2014eea}. We also include the CMB lensing result from the 4-point function (CMBlensing) reported by the Planck collaboration~\citep{Ade:2013aro}. In some cases we replace the results of~\citet{Beutler:2013b} (Beutler2013) with~\citet{Samushia:2013yga} (Samushia2013), \citet{Chuang:2013wga} (Chaung2013) and the BAO only constraints of~\citet{Anderson2.0} (Anderson2013b). In the cases of Beutler2013, Samushia2013 and Chaung2013 we make use of the covariance matrix between the three constraints ($D_V/r_s$, $F_{\rm AP}$ and $f\sigma_8$) presented in the corresponding papers. We also include results using the Planck MCMC chains where the lensing effect on the temperature power spectrum ($A_{\rm L}$-lensing) has been marginalised out (Planck$-A_{\rm L}$). The Planck, Planck$-A_{\rm L}$ and Spergel2013 chains include polarisation results from WMAP (WP). 
}
\scalebox{0.77}{
	\begin{tabular}{lcccc}
		\hline
		dataset(s) & $\sigma_8$ & $\Omega_m$ & \multicolumn{2}{c}{$\sum m_{\nu}\;$[eV]}\\
		& & & $68\%$ c.l. & $95\%$ c.l.\\
		\hline
WMAP9 & $0.706\pm 0.077$ & $0.354^{+0.048}_{-0.078}$ & $< 0.75$ & $< 1.3$\\
WMAP9+CFHTLenS & $0.696^{+0.094}_{-0.071}$ & $0.343^{+0.046}_{-0.078}$ & $< 0.76$ & $< 1.3$\\
WMAP9+Beutler2013 & $0.733\pm 0.038$ & $0.309\pm 0.015$ & $0.36\pm 0.14$ & $0.36\pm 0.28$\\
WMAP9+Beutler2013+CFHTLenS & $0.731\pm 0.026$ & $0.308\pm 0.014$ & $0.37\pm 0.12$ & $0.37\pm 0.24$\\
WMAP9+Beutler2013+GGlensing & $0.725\pm 0.029$ & $0.307\pm 0.014$ & $0.39\pm 0.12$ & $0.39\pm 0.25$\\
WMAP9+Beutler2013+CFHTLenS+GGlensing+BAO & $0.733\pm 0.024$ & $0.303\pm 0.011$ & $0.35\pm 0.10$ & $0.35\pm 0.21$\\
WMAP9+Samushia2013 & $0.746\pm 0.036$ & $0.303\pm 0.013$ & $0.31\pm 0.13$ & $0.31\pm 0.25$\\
WMAP9+Samushia2013+CFHTLenS+GGlensing+BAO & $0.740\pm 0.023$ & $0.2991\pm 0.0097$ & $0.32\pm 0.10$ & $0.32\pm 0.20$\\
WMAP9+Chuang2013 & $0.717\pm 0.046$ & $0.311\pm 0.015$ & $0.42\pm 0.17$ & $0.42\pm 0.35$\\
WMAP9+Chuang2013+CFHTLenS+GGlensing+BAO & $0.728\pm 0.026$ & $0.304\pm 0.011$ & $0.36\pm 0.11$ & $0.36\pm 0.23$\\
WMAP9+Anderson2013 & $0.763^{+0.058}_{-0.040}$ & $0.295\pm 0.011$ & $< 0.31$ & $< 0.54$\\
WMAP9+Anderson2013+BAO & $0.763^{+0.060}_{-0.041}$ & $0.2946\pm 0.0093$ & $< 0.31$ & $< 0.53$\\
WMAP9+Anderson2013+CFHTLenS+GGlensing+BAO & $0.750\pm 0.029$ & $0.2936\pm 0.0097$ & $0.27\pm 0.12$ & $0.27\pm 0.22$\\
\hline
Planck & $0.775^{+0.074}_{-0.031}$ & $0.353^{+0.021}_{-0.058}$ & $< 0.41$ & $< 0.95$\\
Planck+CFHTLenS & $0.745^{+0.083}_{-0.112}$ & $0.332\pm 0.064$ & $< 0.51$ & $< 1.0$\\
Planck+Beutler2013 & $0.791^{+0.034}_{-0.025}$ & $0.320\pm 0.014$ & $0.20\pm 0.13$ & $< 0.40$\\
Planck+Beutler2013+CFHTLenS & $0.760^{+0.026}_{-0.047}$ & $0.314\pm 0.019$ & $0.29\pm 0.13$ & $0.29^{+0.29}_{-0.23}$\\
Planck+Beutler2013+GGlensing & $0.769\pm 0.035$ & $0.316\pm 0.016$ & $0.26\pm 0.13$ & $0.26\pm 0.24$\\
Planck+Beutler2013+CFHTLenS+GGlensing+BAO & $0.759^{+0.025}_{-0.033}$ & $0.306\pm 0.015$ & $0.27\pm 0.12$ & $0.27\pm 0.21$\\
Planck+CMBlensing+Beutler2013+CFHTLenS+GGlensing+BAO & $0.774^{+0.025}_{-0.037}$ & $0.304\pm 0.014$ & $0.24\pm 0.14$ & $0.24\pm 0.20$\\
Planck+Samushia2013 & $0.800^{+0.029}_{-0.023}$ & $0.315\pm 0.013$ & $0.161^{+0.068}_{-0.139}$ & $< 0.33$\\
Planck+Samushia2013+CFHTLenS+GGlensing+BAO & $0.765\pm 0.031$ & $0.303\pm 0.014$ & $0.243^{+0.132}_{-0.088}$ & $0.24\pm 0.19$\\
Planck+Chuang2013 & $0.797^{+0.038}_{-0.026}$ & $0.319\pm 0.014$ & $< 0.23$ & $< 0.40$\\
Planck+Chuang2013+CFHTLenS+GGlensing+BAO & $0.759^{+0.027}_{-0.037}$ & $0.306\pm 0.015$ & $0.27\pm 0.12$ & $0.27\pm 0.22$\\
Planck+Anderson2013 & $0.821^{+0.023}_{-0.012}$ & $0.304\pm 0.010$ & $< 0.10$ & $< 0.22$\\
Planck+Anderson2013+BAO & $0.821^{+0.022}_{-0.013}$ & $0.3020\pm 0.0084$ & $< 0.09$ & $< 0.21$\\
Planck+Anderson2013+CFHTLenS+GGlensing+BAO & $0.782^{+0.029}_{-0.048}$ & $0.296^{+0.010}_{-0.015}$ & $0.17\pm 0.12$ & $< 0.33$\\
Planck+CMBlensing+Anderson2013+CFHTLenS+GGlensing+BAO & $0.794^{+0.025}_{-0.032}$ & $0.294\pm 0.012$ & $0.15^{+0.15}_{-0.12}$ & $< 0.30$\\
Planck+CMBlensing & $0.746^{+0.086}_{-0.038}$ & $0.373^{+0.048}_{-0.077}$ & $< 0.62$ & $< 1.1$\\
\hline
Planck$-A_{\rm L}$ & $0.716^{+0.092}_{-0.064}$ & $0.356^{+0.043}_{-0.065}$ & $< 0.71$ & $< 1.2$\\
Planck$-A_{\rm L}$+CFHTLenS & $0.694^{+0.099}_{-0.079}$ & $0.351^{+0.048}_{-0.076}$ & $0.62^{+0.36}_{-0.50}$ & $< 1.3$\\
Planck$-A_{\rm L}$+Beutler2013 & $0.746\pm 0.035$ & $0.316\pm 0.015$ & $0.34\pm 0.14$ & $0.34\pm 0.26$\\
Planck$-A_{\rm L}$+Beutler2013+CFHTLenS & $0.733\pm 0.027$ & $0.314^{+0.013}_{-0.018}$ & $0.38\pm 0.11$ & $0.38\pm 0.24$\\
Planck$-A_{\rm L}$+Beutler2013+GGlensing & $0.733\pm 0.031$ & $0.314^{+0.013}_{-0.017}$ & $0.38\pm 0.12$ & $0.38\pm 0.25$\\
Planck$-A_{\rm L}$+Beutler2013+CFHTLenS+GGlensing+BAO & $0.736\pm 0.024$ & $0.307\pm 0.011$ & $0.36\pm 0.10$ & $0.36\pm 0.21$\\
Planck$-A_{\rm L}$+CMBlensing+Beutler2013+CFHTLenS+GGlensing+BAO & $0.731^{+0.030}_{-0.040}$ & $0.309\pm 0.015$ & $0.38^{+0.12}_{-0.17}$ & $0.38\pm 0.20$\\
Planck$-A_{\rm L}$+Samushia2013 & $0.759\pm 0.035$ & $0.310\pm 0.013$ & $0.28\pm 0.12$ & $0.28\pm 0.23$\\
Planck$-A_{\rm L}$+Samushia2013+CFHTLenS+GGlensing+BAO & $0.743\pm 0.024$ & $0.303\pm 0.011$ & $0.324\pm 0.099$ & $0.32\pm 0.19$\\
Planck$-A_{\rm L}$+Chuang2013 & $0.737\pm 0.042$ & $0.318\pm 0.016$ & $0.38^{+0.15}_{-0.19}$ & $0.38\pm 0.32$\\
Planck$-A_{\rm L}$+Chuang2013+CFHTLenS+GGlensing+BAO & $0.730\pm 0.028$ & $0.309\pm 0.012$ & $0.38\pm 0.11$ & $0.38\pm 0.22$\\
Planck$-A_{\rm L}$+Anderson2013 & $0.784^{+0.046}_{-0.026}$ & $0.299^{+0.010}_{-0.013}$ & $< 0.23$ & $< 0.43$\\
Planck$-A_{\rm L}$+Anderson2013+BAO & $0.785^{+0.046}_{-0.029}$ & $0.2985\pm 0.0094$ & $< 0.23$ & $< 0.42$\\
Planck$-A_{\rm L}$+Anderson2013+CFHTLenS+GGlensing+BAO & $0.755\pm 0.028$ & $0.297\pm 0.010$ & $0.27\pm 0.11$ & $0.27\pm 0.21$\\
Planck$-A_{\rm L}$+CMBlensing+Anderson2013+CFHTLenS+GGlensing+BAO & $0.747\pm 0.036$ & $0.300\pm 0.014$ & $0.30\pm 0.15$ & $0.30\pm 0.24$\\
Planck$-A_{\rm L}$+CMBlensing & $0.658^{+0.036}_{-0.062}$ & $0.400^{+0.066}_{-0.051}$ & $0.86^{+0.35}_{-0.28}$ & $0.86\pm 0.67$\\
\hline
Spergel2013 & $0.761^{+0.074}_{-0.033}$ & $0.343^{+0.024}_{-0.056}$ & $< 0.44$ & $< 0.92$\\
Spergel2013+CFHTLenS & $0.741^{+0.077}_{-0.058}$ & $0.329^{+0.034}_{-0.058}$ & $< 0.50$ & $< 0.99$\\
Spergel2013+Beutler2013 & $0.774\pm 0.029$ & $0.317\pm 0.015$ & $0.24\pm 0.12$ & $0.24\pm 0.22$\\
Spergel2013+Beutler2013+CFHTLenS & $0.753^{+0.025}_{-0.032}$ & $0.313\pm 0.016$ & $0.30\pm 0.11$ & $0.30\pm 0.23$\\
Spergel2013+Beutler2013+GGlensing & $0.758^{+0.029}_{-0.037}$ & $0.314\pm 0.015$ & $0.29\pm 0.12$ & $0.29\pm 0.23$\\
Spergel2013+Beutler2013+CFHTLenS+GGlensing+BAO & $0.754^{+0.024}_{-0.033}$ & $0.306\pm 0.011$ & $0.29\pm 0.10$ & $0.29\pm 0.20$\\
Spergel2013+Samushia2013 & $0.784\pm 0.028$ & $0.312\pm 0.013$ & $0.201^{+0.091}_{-0.113}$ & $< 0.38$\\
Spergel2013+Samushia2013+CFHTLenS+GGlensing+BAO & $0.760\pm 0.024$ & $0.303\pm 0.011$ & $0.26\pm 0.10$ & $0.26\pm 0.19$\\
Spergel2013+Chuang2013 & $0.777\pm 0.034$ & $0.317\pm 0.016$ & $0.24^{+0.11}_{-0.15}$ & $< 0.47$\\
Spergel2013+Chuang2013+CFHTLenS+GGlensing+BAO & $0.752^{+0.025}_{-0.032}$ & $0.307\pm 0.011$ & $0.29\pm 0.11$ & $0.29\pm 0.22$\\
Spergel2013+Anderson2013 & $0.807^{+0.028}_{-0.016}$ & $0.300\pm 0.011$ & $< 0.14$ & $< 0.27$\\
Spergel2013+Anderson2013+BAO & $0.808^{+0.027}_{-0.015}$ & $0.2992\pm 0.0086$ & $< 0.13$ & $< 0.26$\\
Spergel2013+Anderson2013+CFHTLenS+GGlensing+BAO & $0.776\pm 0.027$ & $0.296\pm 0.010$ & $0.191^{+0.098}_{-0.122}$ & $< 0.36$\\
\hline
	  \end{tabular}
	  }
	  \label{tab:para2}
\end{center}
\end{table*}

Knowing that our CMASS measurements and the weak lensing constraints are quite insensitive to the fiducial neutrino mass we now combine these constraints with CMB data from WMAP9\footnote{\url{http://lambda.gsfc.nasa.gov/product/map/current/}} and Planck\footnote{\url{http://irsa.ipac.caltech.edu/data/Planck/release_1/ancillary-data/}}. We importance sample the CMB MCMC chains, where $\sum m_{\nu}$ is varied freely. Importance sampling means that we adjust the weight of each chain element in the original CMB chain by the likelihood, $\mathcal{L}\sim \exp(-\chi^2/2)$, of some external dataset, according to
\begin{equation}
w_{\rm new} = w_{\rm CMB}\times \mathcal{L}.
\end{equation}
The distribution of these new weights reflects the combined likelihood.

To provide constraints on some combination of $\sigma_8$ and $\Omega_m$ we adopted the CMASS $f\sigma_8$ measurement of~\citet{Beutler:2013b}, the CFHTLenS constraint reported in~\citet{Kilbinger:2012qz},
\begin{equation}
\sigma_8\left(\frac{\Omega_m}{0.27}\right)^{0.6} = 0.79\pm0.03
\label{eq:CFHTLS}
\end{equation}
and the galaxy-galaxy lensing result reported in~\citet{Mandelbaum:2012ay},
\begin{equation}
\sigma_8\left(\frac{\Omega_m}{0.25}\right)^{0.57} = 0.80\pm0.05.
\label{eq:GGlensing}
\end{equation}
We prefer to use the full degeneracy lines (eq.~\ref{eq:CFHTLS} and~\ref{eq:GGlensing}) from these lensing results instead of the actual likelihoods. This approach is more conservative and does not affect our final results given the power of the BAO scale to break the degeneracy between $\Omega_m$ and $\sigma_8$.
We occasionally also include the BAO constraint of 6dFGS~\citep{Beutler:2011hx} and LOWZ~\citep{Anderson2.0,Tojeiro:2014eea}. In section~\ref{sec:dis} we will discuss the significance with which these datasets prefer $\Lambda$CDM$+\sum m_{\nu}$ over $\Lambda$CDM.

\begin{figure*}
\begin{center}
\epsfig{file=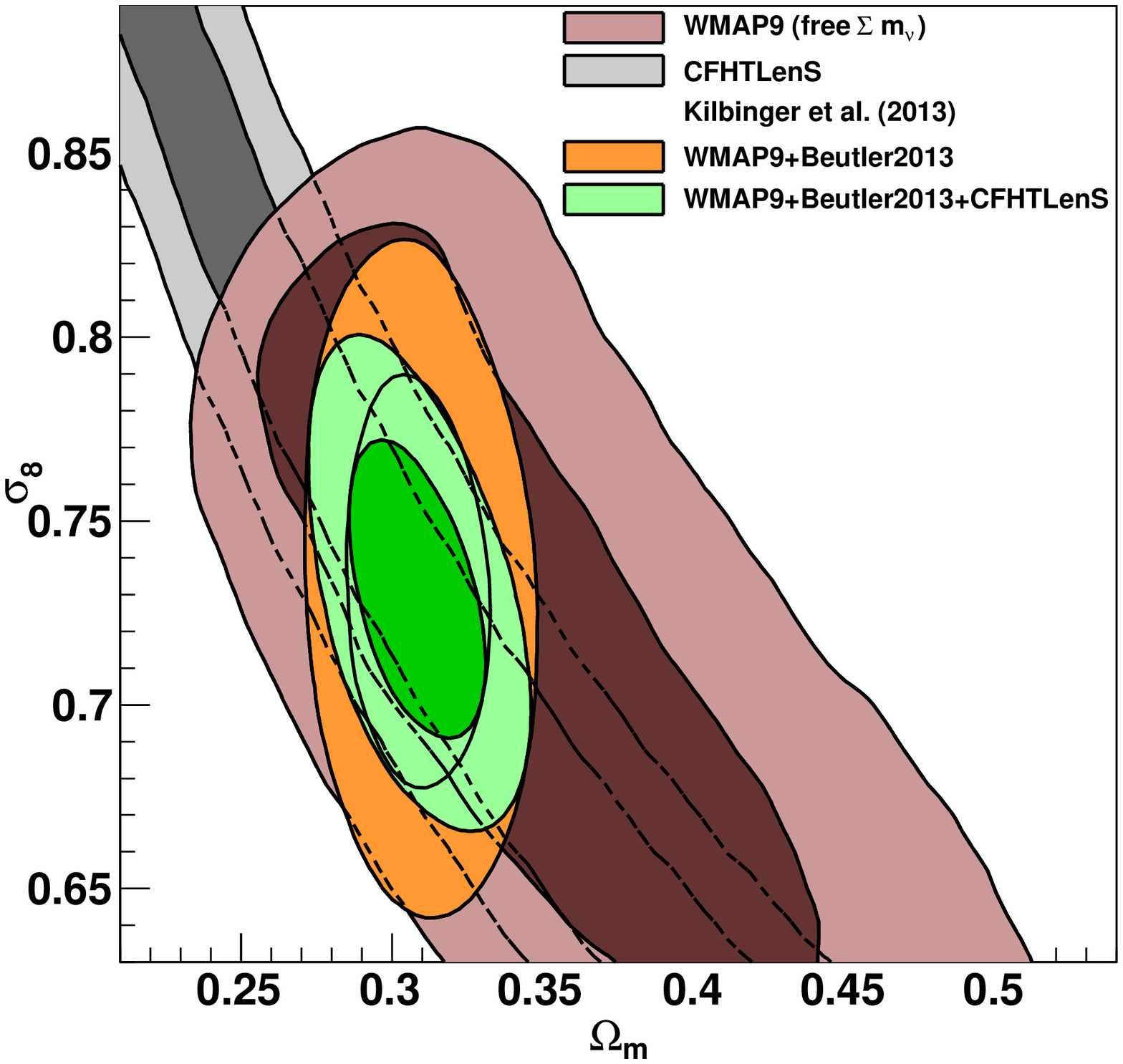,width=8.8cm}
\epsfig{file=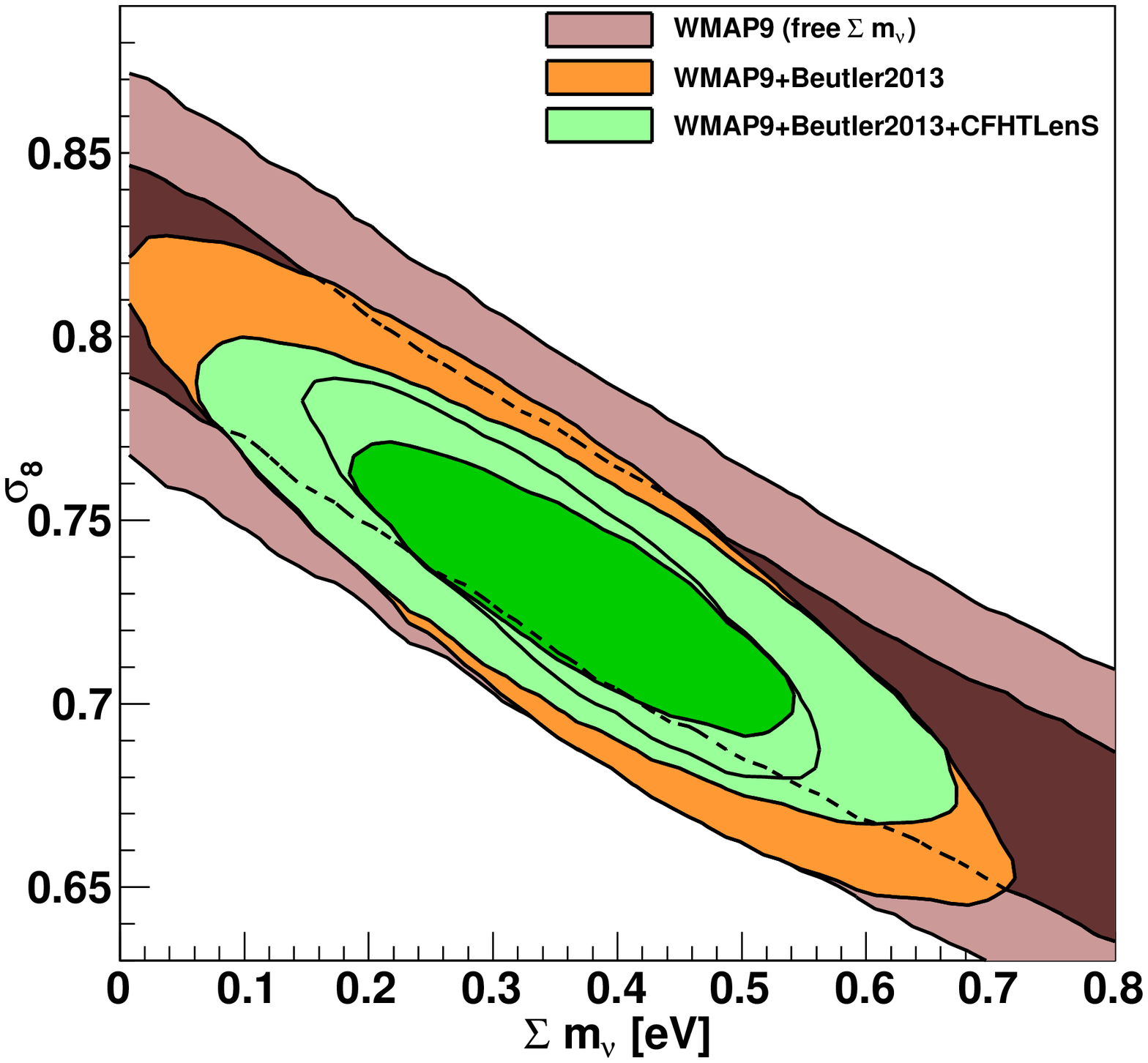,width=8.8cm}
\caption{Two-dimensional likelihood for $\Omega_m$-$\sigma_8$ (left) and $\sum m_{\nu}$-$\sigma_8$ (right) when combining the WMAP9 MCMC chain within $\Lambda$CDM and free $\sum m_{\nu}$ with different low redshift growth of structure constraints. The orange contours show WMAP9+Beutler2013, where Beutler2013 stands for the constraints on $D_V/r_s$, $F_{\rm AP}$ and $f\sigma_8$ reported in~\citet{Beutler:2013b}. The green contours show WMAP9+Beutler2013+CFHTLenS. The results are summarised in Table~\ref{tab:para2}.}
\label{fig:mnu}
\end{center}
\end{figure*}

\subsection{Combining with WMAP9}

\begin{figure}
\begin{center}
\epsfig{file=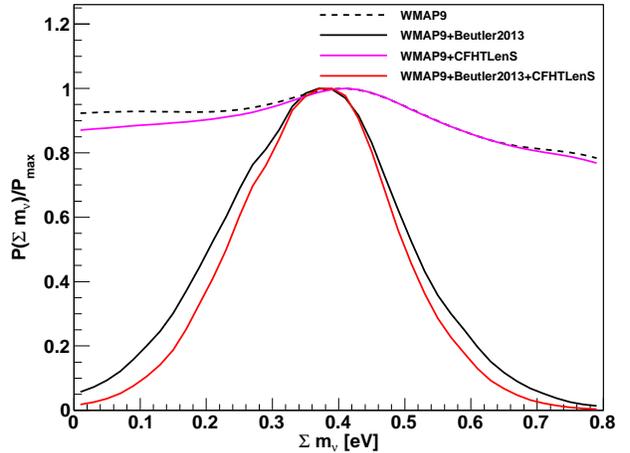,width=8.8cm}
\caption{One-dimensional likelihood distribution for $\sum m_{\nu}$ using WMAP9 combined with different datasets. Beutler2013 stands for the $D_V/r_s$, $F_{\rm AP}$ and $f\sigma_8$ constraints of~\citet{Beutler:2013b}, while CFHTLenS represents the constraint of eq.~\ref{eq:CFHTLS} reported in~\citet{Kilbinger:2012qz} (see also Table~\ref{tab:para2}).}
\label{fig:1Dprop}
\end{center}
\end{figure}

We start with importance sampling the WMAP9 chains\footnote{\url{htp://lambda.gsfc.nasa.gov/product/map/current/}}. To illustrate the constraining power of the different datasets we use them one by one, before combining them. These results are summarised in Table~\ref{tab:para2} and shown in Figure~\ref{fig:mnu} and Figure~\ref{fig:1Dprop}. Using WMAP9 alone already allows a constraint on $\sum m_{\nu}$ of $< 1.3\,$eV with $95\%$ confidence level. If we add BAO constraints, where we use the isotropic CMASS constraint after density field reconstruction from~\citet{Anderson2.0} (Anderson2013b), we obtain an upper limit of $< 0.54\,$eV with $95\%$ confidence level. Therefore adding current BAO information already improves the constraint by more than a factor of two. The CMB combined with BAO constrain the neutrino mass purely from its effect on the geometry of the Universe, with the BAO particularly helping to break the degeneracy between $\sum m_{\nu}$ and $H_{0}$ (see e.g.~\citealt{Hou:2012xq}).

Next we use the lensing results from CFHTLenS~\citep{Kilbinger:2012qz} and galaxy-galaxy lensing~\citep{Mandelbaum:2012ay} (GGlensing). The degeneracy lines in these lensing results are similar to the degeneracy in the CMB and therefore combining WMAP9 with only weak lensing measurements does not improve the neutrino mass constraint significantly (see Table~\ref{tab:para2} for details). The true power of the lensing datasets arises with the addition of the BAO information. The BAO constraints basically fix $\Omega_m$, which, combined with weak lensing allows tight constraints on $\sigma_8$. These $\sigma_8$ constraints are almost a factor of two better than CMB+BAO and improve the neutrino mass constraint significantly. The same arguments hold for the $f\sigma_8$ constraint from galaxy surveys. Since one can obtain both, the BAO scale measurement and the growth of structure measurement from galaxy surveys, we can combine just two datasets, WMAP9 and CMASS, to obtain tight constraints on the neutrino mass. Combining the CMASS-DR11 results for $D_V/r_s$, $F_{\rm AP}$ and $f\sigma_8$ reported in eq.~68 and 70 of~\citet{Beutler:2013b} with WMAP9 produces $\sum m_{\nu} = 0.36\pm 0.14\,$eV, which represents a $2.6\sigma$ preference for neutrino mass. Adding CFHTLenS further improves this constraint to $\sum m_{\nu} = 0.37\pm 0.12\,$eV, which represents a $3.1\sigma$ detection. Replacing CFHTLenS with galaxy-galaxy lensing from~\citet{Mandelbaum:2012ay} (GGlensing) leads to basically the same result. Combining all datasets, including further BAO constraints from 6dFGS and LOWZ, yields $\sum m_{\nu} = 0.35\pm 0.10\,$eV ($3.3\sigma$). The final result does not change significantly if we use the constraints of~\citet{Samushia:2013yga} (Samushia2013b), \citet{Chuang:2013wga} (Chuang2013) or the BAO-only constraints of~\citet{Anderson2.0} (Anderson2013b) instead of~\citet{Beutler:2013b}. 

\subsection{Combining with Planck}

\begin{figure*}
\begin{center}
\epsfig{file=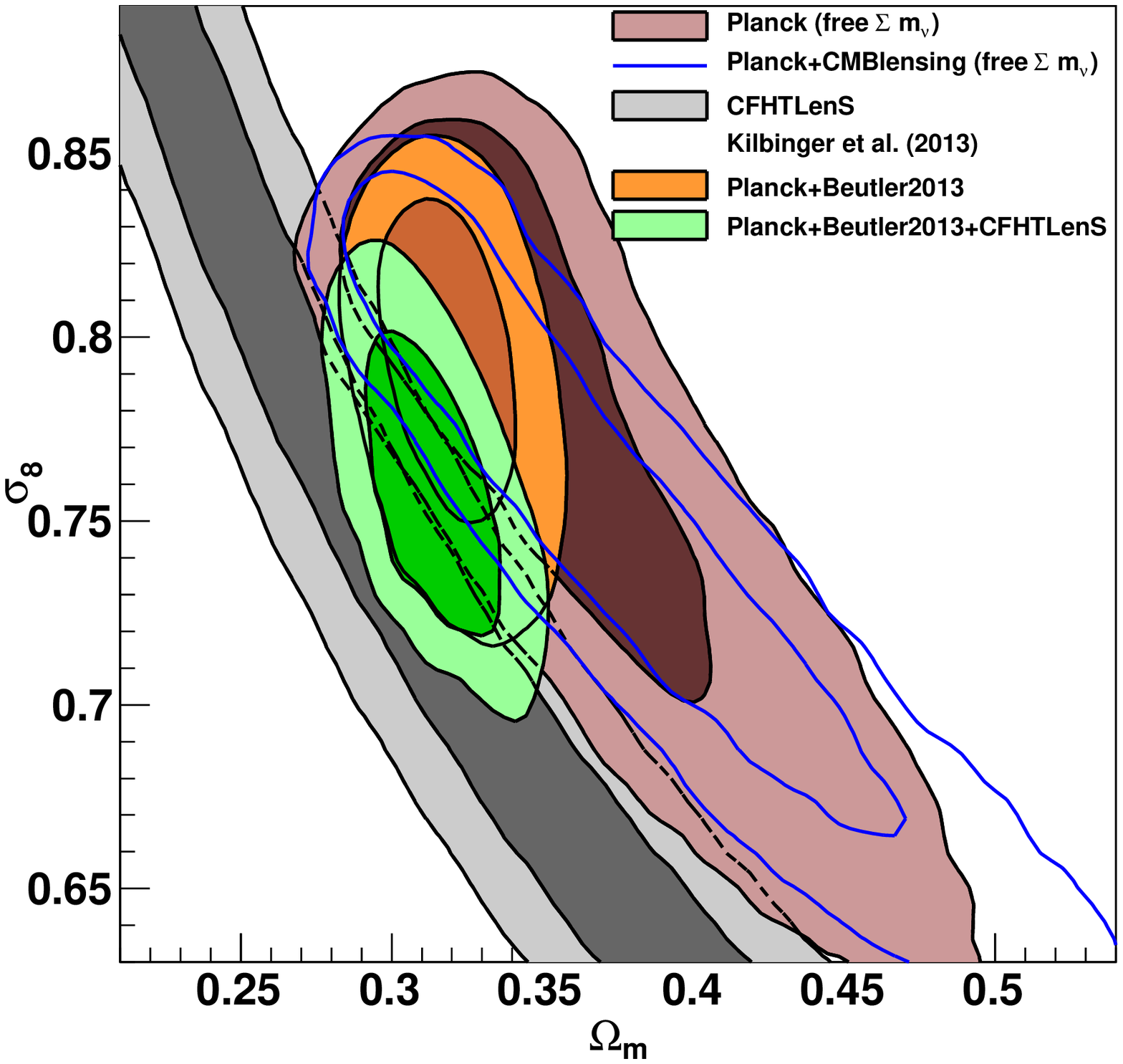,width=8.8cm}
\epsfig{file=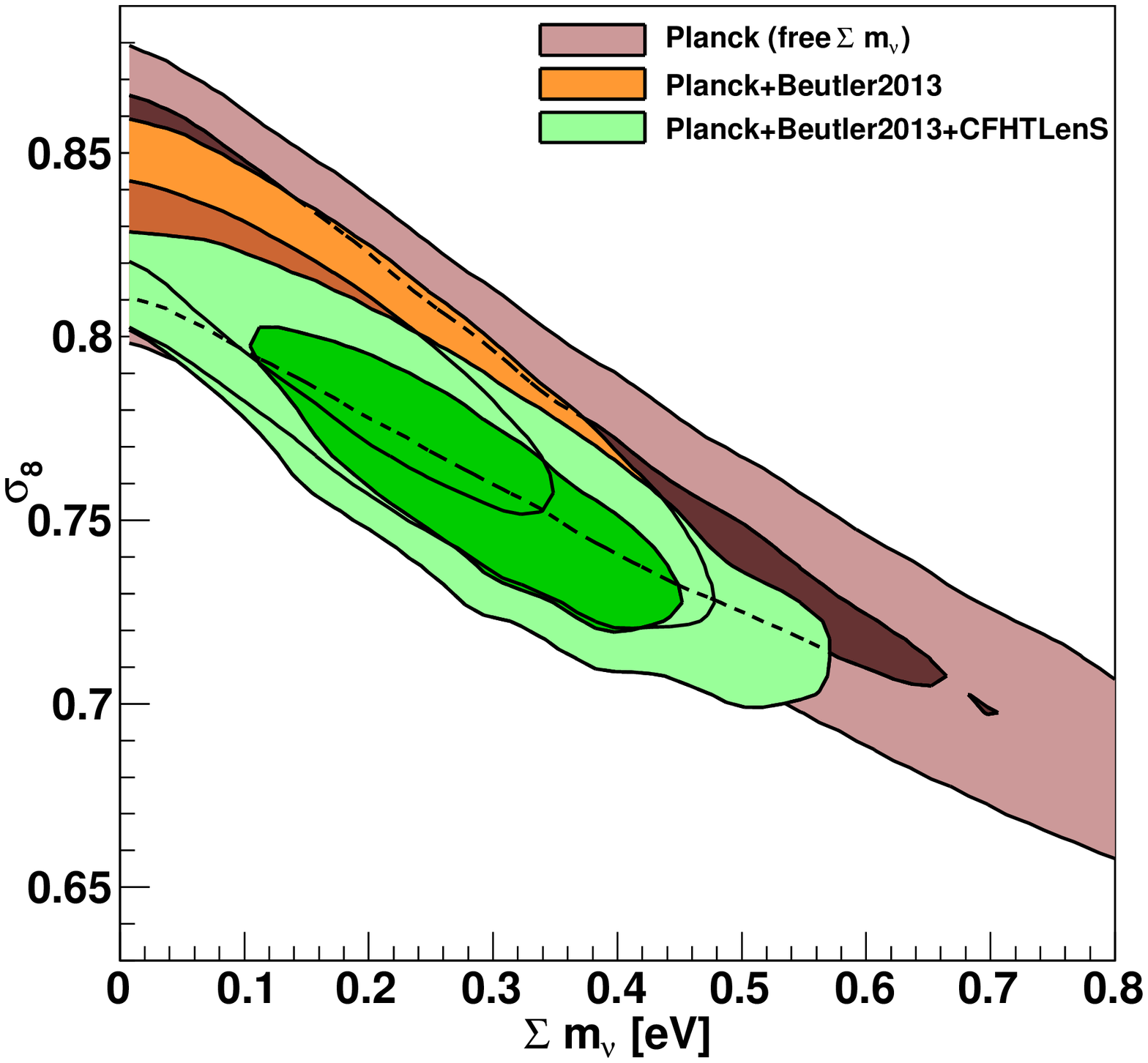,width=8.8cm}
\epsfig{file=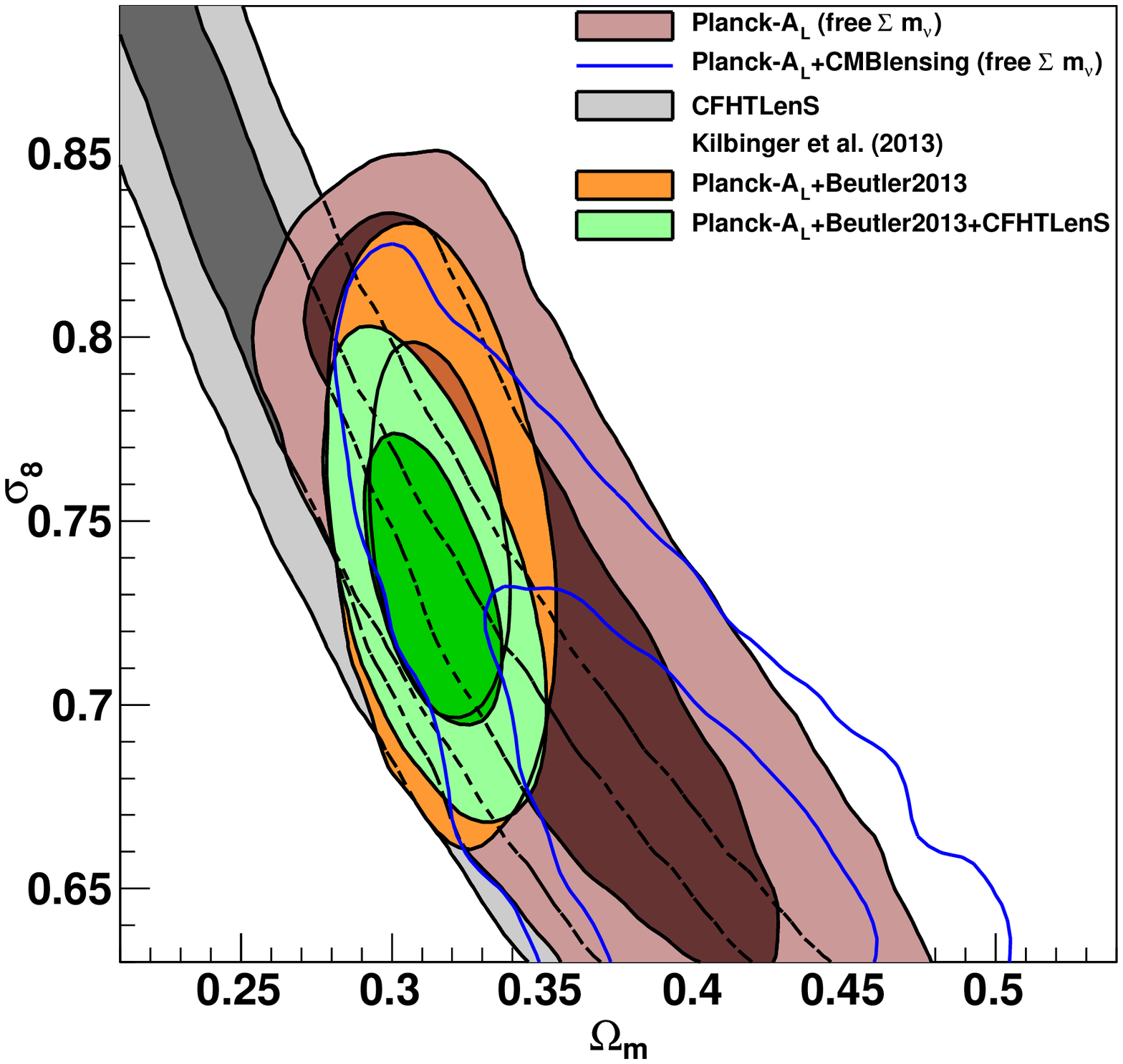,width=8.8cm}
\epsfig{file=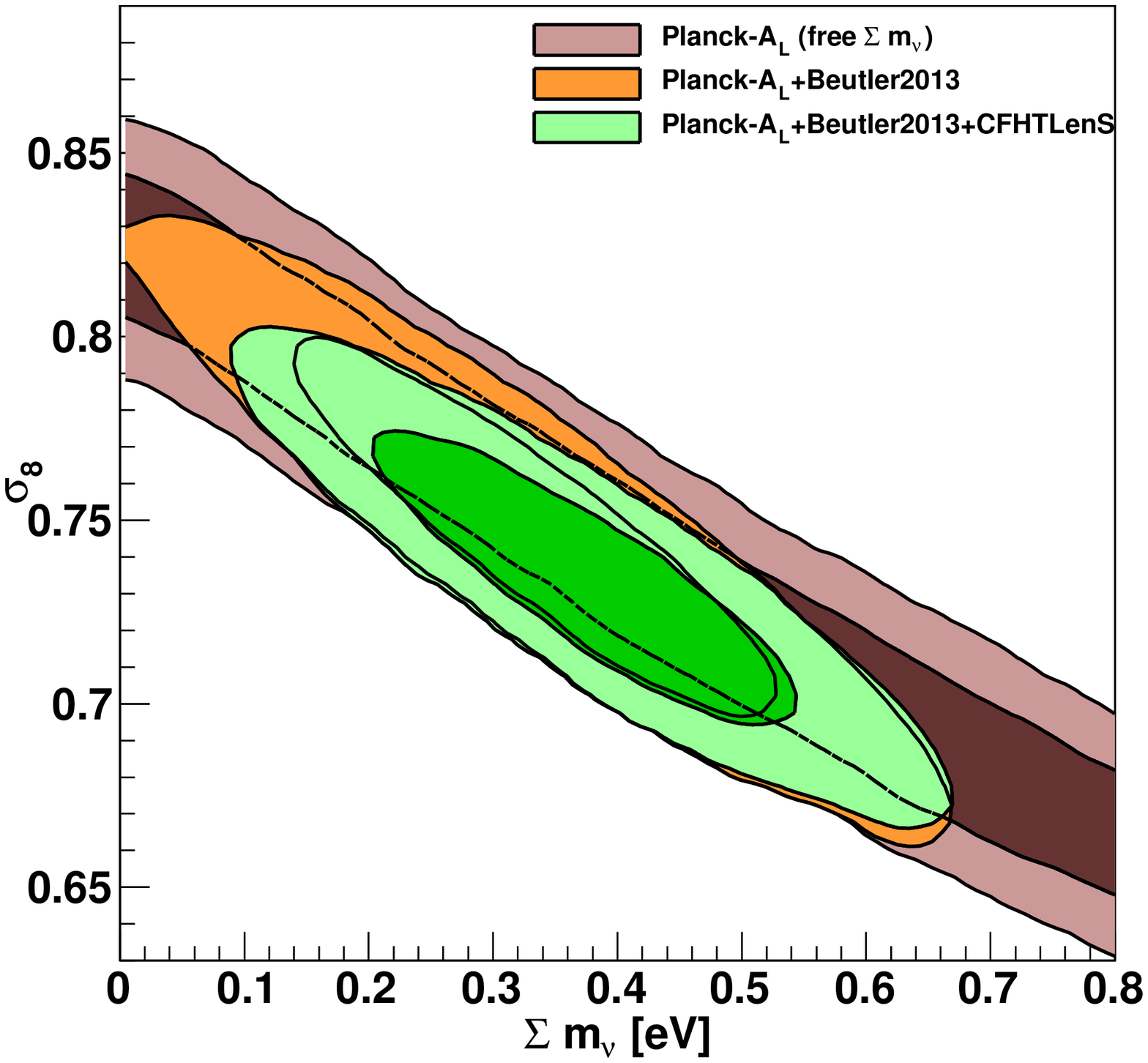,width=8.8cm}
\caption{Two-dimensional likelihood for $\Omega_m$-$\sigma_8$ (left) and $\sum m_{\nu}$-$\sigma_8$ (right) when combining Planck MCMC chains within $\Lambda$CDM and free $\sum m_{\nu}$ with different low redshift growth of structure constraints. We show the main Planck results in the two plots on the top. The two bottom plots show the results where we used a Planck MCMC chain with the $A_{\rm L}$-lensing signal marginalised out. The orange contours show Planck combined with the $D_V/r_s$, $F_{\rm AP}$ and $f\sigma_8$ constraints of~\citet{Beutler:2013b}. The green contours additionally include CFHTLenS. The blue contours show Planck and Planck$-A_{\rm L}$ combined with CMB lensing from the $4$-point function (top left and bottom left, respectively). The results are summarised in Table~\ref{tab:para2}.}
\label{fig:mnu2}
\end{center}
\end{figure*}

We now replace the WMAP9 dataset with Planck\footnote{\url{http://irsa.ipac.caltech.edu/data/Planck/release_1/
ancillary-data/}} and repeat the exercise of the last section. The results are summarised in Table~\ref{tab:para2} and shown in Figure~\ref{fig:mnu2} (upper two plots) and Figure~\ref{fig:1Dprop2} (top). Combining Planck with the results of~\citet{Beutler:2013b} yields a mild preference for neutrino mass of $\sum m_{\nu} = 0.20\pm0.13\,$eV ($68\%$ c.l.) or an upper limit of $< 0.40\,$eV with $95\%$ confidence level. 
Including CFHTLenS produces $\sum m_{\nu} = 0.29\pm 0.13\,$eV. Adding galaxy-galaxy lensing or further BAO measurements results in a $2.3\sigma$ preference of neutrino mass yielding $\sum m_{\nu} = 0.27\pm 0.12\,$eV. When replacing the result of~\citet{Beutler:2013b} with~\citet{Samushia:2013yga},~\citet{Chuang:2013wga} or~\citet{Anderson2.0} we find consistent results. Note that since Planck is in tension with some of the other datasets, importance sampling does rely on fairly low density regions in the Planck chains. We conclude that these results cannot be used to claim a significant detection of the neutrino mass.

\subsection{Combining with Planck marginalised over $A_{\rm L}$}

Table~\ref{tab:para2} also includes results using Planck MCMC chains where the lensing contribution to the temperature power spectrum has been marginalised out. We denote this chain Planck$-A_{\rm L}$ since $A_{\rm L}$ is the parameter used to mimic the lensing effect on the CMB temperature power spectrum (smoothing of the higher order peaks). 
One must keep in mind, however, that $A_{\rm L}$ is not a physical parameter, but only a way to remove the lensing effect from the CMB power spectrum data. To avoid confusion, from now on we will designate the lensing contribution to the temperature power spectrum as $A_{\rm L}$-lensing and the lensing signal in the 4-point function as CMB lensing (or CMBlensing in Table~\ref{tab:para2}). The WMAP dataset is not sensitive to gravitational lensing, because this effect is only significant at large multipoles.

The Planck collaboration reports some anomalies with respect to the $A_{\rm L}$-lensing contribution. When including the parameter $A_{\rm L}$ in the fit, Planck reports $A_{\rm L} = 1.29\pm 0.13$ (Planck+WP)~\citep{Ade:2013zuv}, which is $2\sigma$ from the expected value of $1$, while the lensing effect in the 4-point function produces $A^{\phi\phi}_{\rm L} = 0.99\pm 0.05$~\citep{Ade:2013aro}. Thus the $A_{\rm L}$-lensing contribution is in (small) tension with the overall Planck results and with the $4$-point function lensing of Planck. The CMB lensing of Planck favours larger neutrino masses compared to the rest of Planck, which therefore leads to a weakening of the neutrino mass constraints when CMB lensing is included (see~\citealt{Ade:2013zuv} section 6.3.1).
We show these results in Table~\ref{tab:para2}, where Planck+WP gives $\sum m_{\nu} < 0.95\,$eV, while Planck+WP+CMBlensing yields $\sum m_{\nu} < 1.1\,$eV. While the neutrino mass constraints improve if ACT~\citep{Das:2013zf} and SPT~\citep{Keisler:2011aw} data (highL) are included, it does not relieve the tensions with $A_{\rm L}$-lensing. Because of the points mention above it is interesting to investigate what happens with the Planck data when the $A_{\rm L}$-lensing signal is excluded.

Excluding the $A_{\rm L}$-lensing contribution significantly degrades the constraints on the neutrino mass from Planck alone, since these constraints are dominated by the $A_{\rm L}$-lensing effect. This, however, has little effect on our analysis, since we can break the $\sum m_{\nu}$-$\sigma_8$ degeneracy more efficiently with the low-redshift datasets. Another effect of marginalising over $A_{\rm L}$ is much more significant. Marginalising over $A_{\rm L}$ leads to $1\sigma$ shifts in $\Omega_m$ and $\sigma_8$. Within $\Lambda$CDM including the default value of $\sum m_{\nu} = 0.06\,$eV, the Planck team found for $\Omega_m$:
\begin{align}
\Omega_m &= 0.315^{+0.016}_{-0.018}\hspace{0.8cm}(\text{Planck+WP}),\\
\Omega_m &= 0.295^{+0.017}_{-0.020}\hspace{0.8cm}(\text{Planck}-A_{\rm L}\text{+WP})
\end{align}
and for $\sigma_8$:
\begin{align}
\sigma_8 &= 0.829\pm0.012\hspace{0.8cm}(\text{Planck+WP}),\\
\sigma_8 &= 0.814\pm0.014\hspace{0.8cm}(\text{Planck}-A_{\rm L}\text{+WP}).
\end{align}
These shifts bring Planck in much better agreement with WMAP9. Since we still have a high value of $\sigma_8$ compared to the growth of structure measurements, we still have a preference for a neutrino mass, similar to the results in WMAP9. 

Combining Planck$-A_{\rm L}$ with the results of~\citet{Beutler:2013b} produces $\sum m_{\nu} = 0.34\pm 0.14\,$eV, in excellent agreement with the result obtained when combining with WMAP9. 
Including CFHTLenS yields $\sum m_{\nu} = 0.38\pm 0.11\,$eV, and adding galaxy-galaxy lensing and further BAO constraints improves this detection to $\sum m_{\nu} = 0.36\pm 0.10\,$eV ($3.4\sigma$). This detection is robust against various dataset variations as shown in Table~\ref{tab:para2}. The results are also presented in Figure~\ref{fig:mnu2} (lower two plots) and Figure~\ref{fig:1Dprop2} (middle). 

\subsection{Combining with the Planck re-analysis of~\citet{Spergel:2013rxa}}

\citet{Spergel:2013rxa} re-analysed the Planck data with a different treatment for foreground cleaning, which has a notable effect on the $217\,$GHz spectra. From now on we will call this analysis Spergel2013. Their result shows $\sim 1\sigma$ shifts in $\sigma_8$ and $\Omega_m$ towards smaller values. Similar shifts caused by different foreground removal techniques have been reported by the Planck collaboration~\citep{Ade:2013hta}. These changes in $\Omega_m$ and $\sigma_8$ are smaller, but similar to the shifts we found by excluding the $A_{\rm L}$-lensing contribution. We saw that such shifts can significantly alter the constraints on $\sum m_{\nu}$. 

\begin{figure}
\begin{center}
\epsfig{file=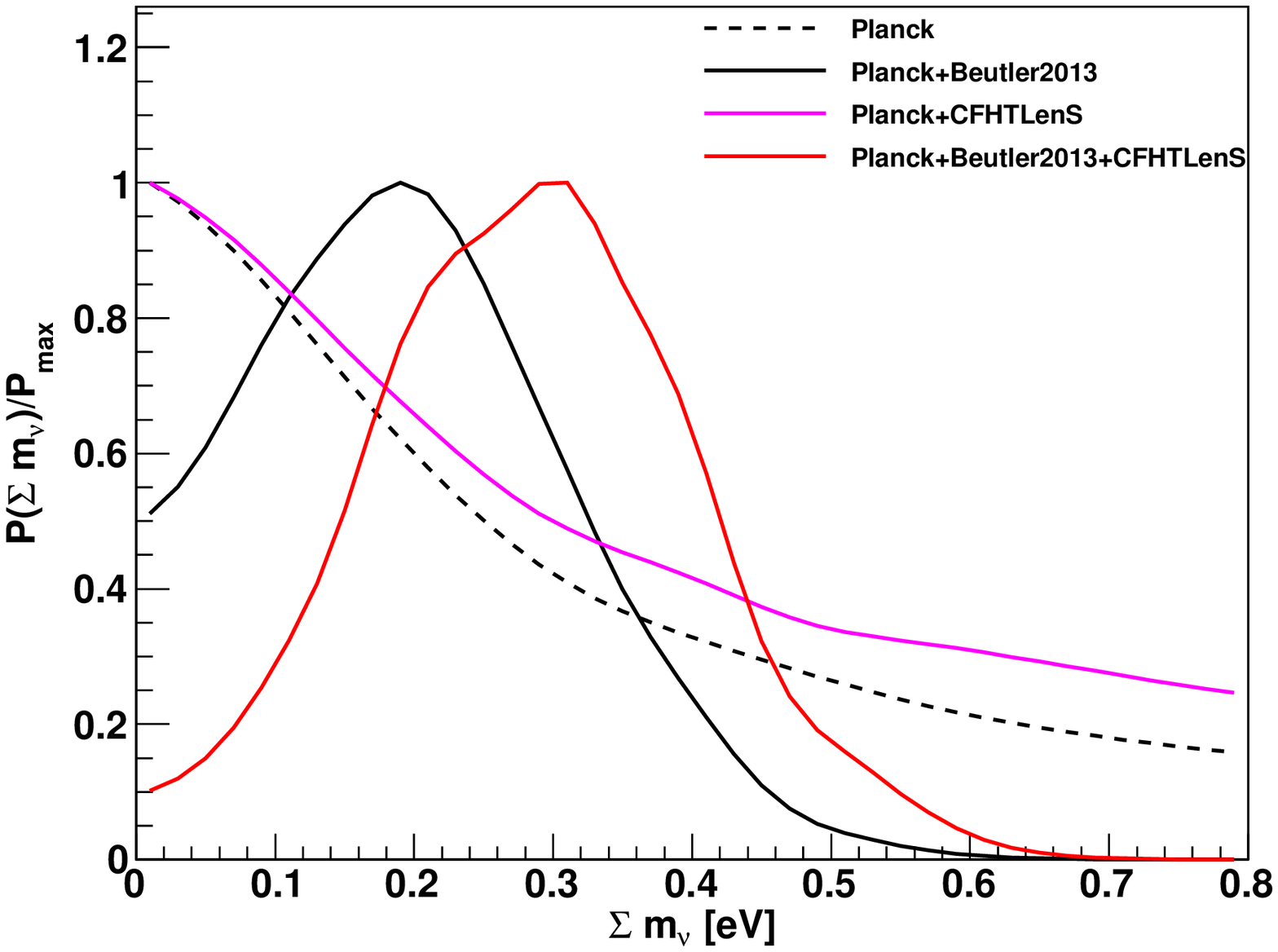,width=8.8cm}\\
\epsfig{file=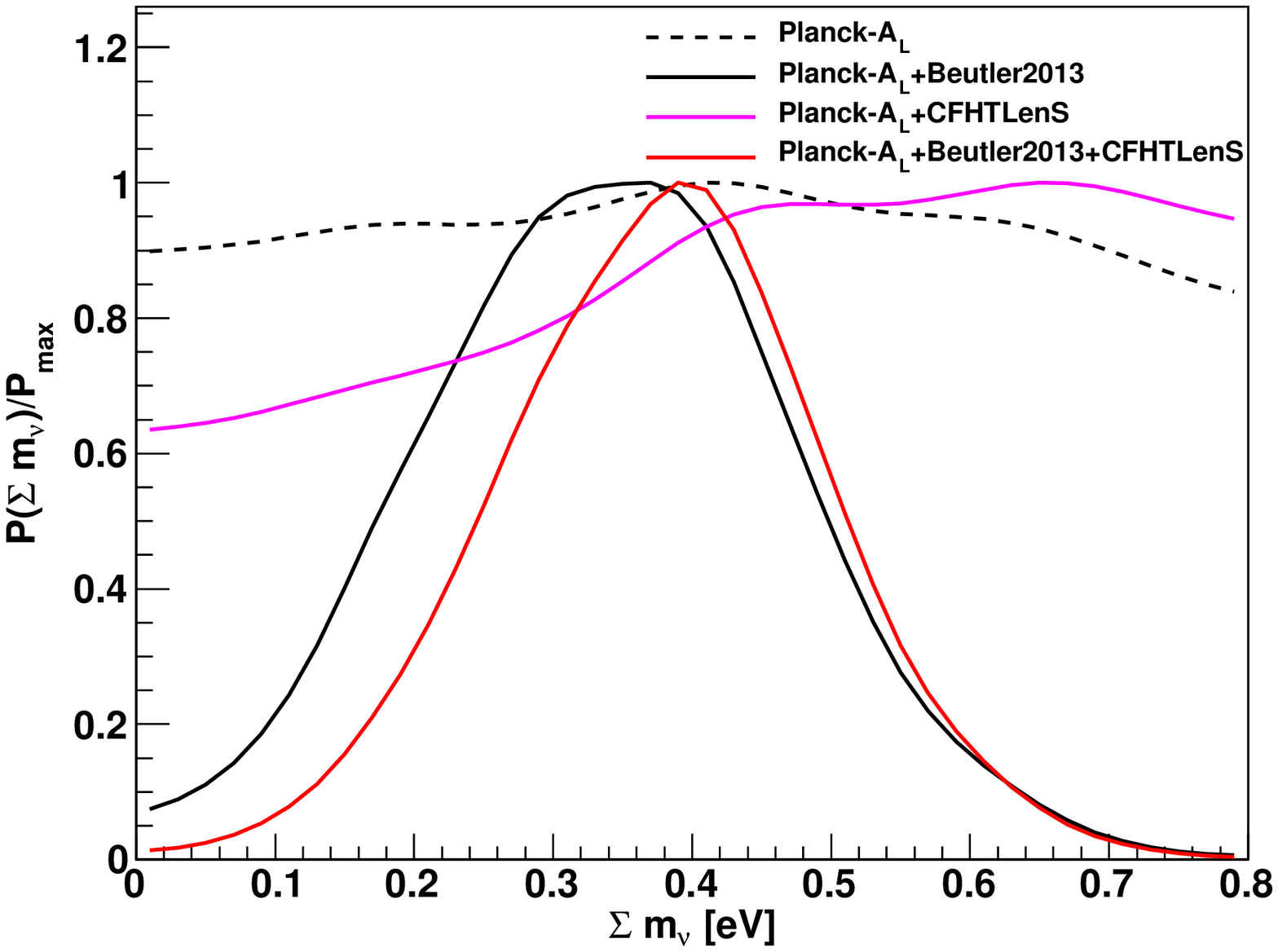,width=8.8cm}\\
\epsfig{file=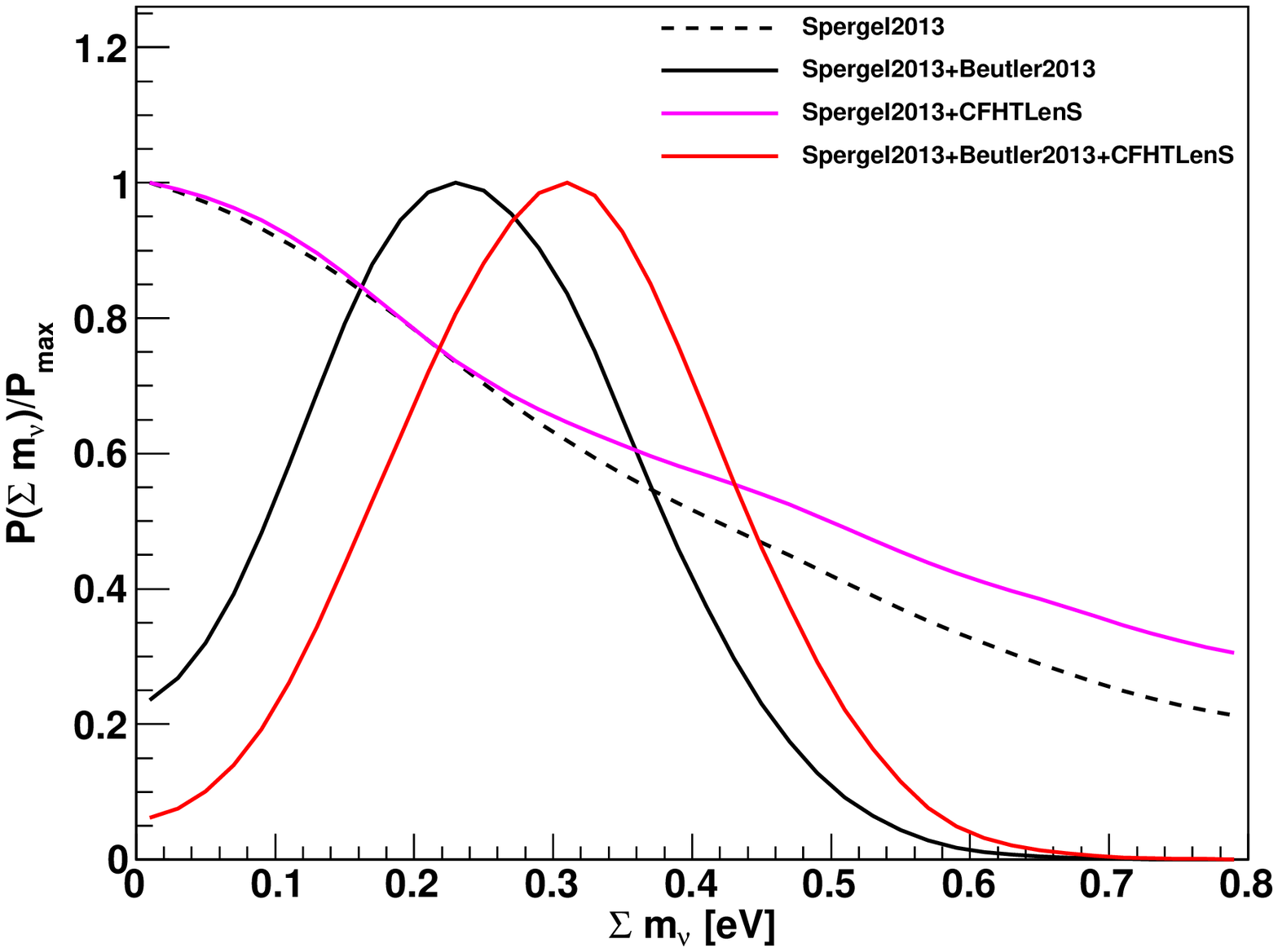,width=8.8cm}
\caption{One-dimensional likelihood distribution for $\sum m_{\nu}$ when combining Planck with different datasets. We show the results for the main Planck dataset (top), the Planck dataset without the $A_{\rm L}$-lensing signal (middle) and the Planck re-analysis of~~\citet{Spergel:2013rxa} (bottom). Beutler2013 stands for the $D_V/r_s$, $F_{\rm AP}$ and $f\sigma_8$ constraints of~\citet{Beutler:2013b}, while CFHTLenS represents the constraint reported in~\citet{Kilbinger:2012qz}.}
\label{fig:1Dprop2}
\end{center}
\end{figure}

\begin{figure*}
\begin{center}
\epsfig{file=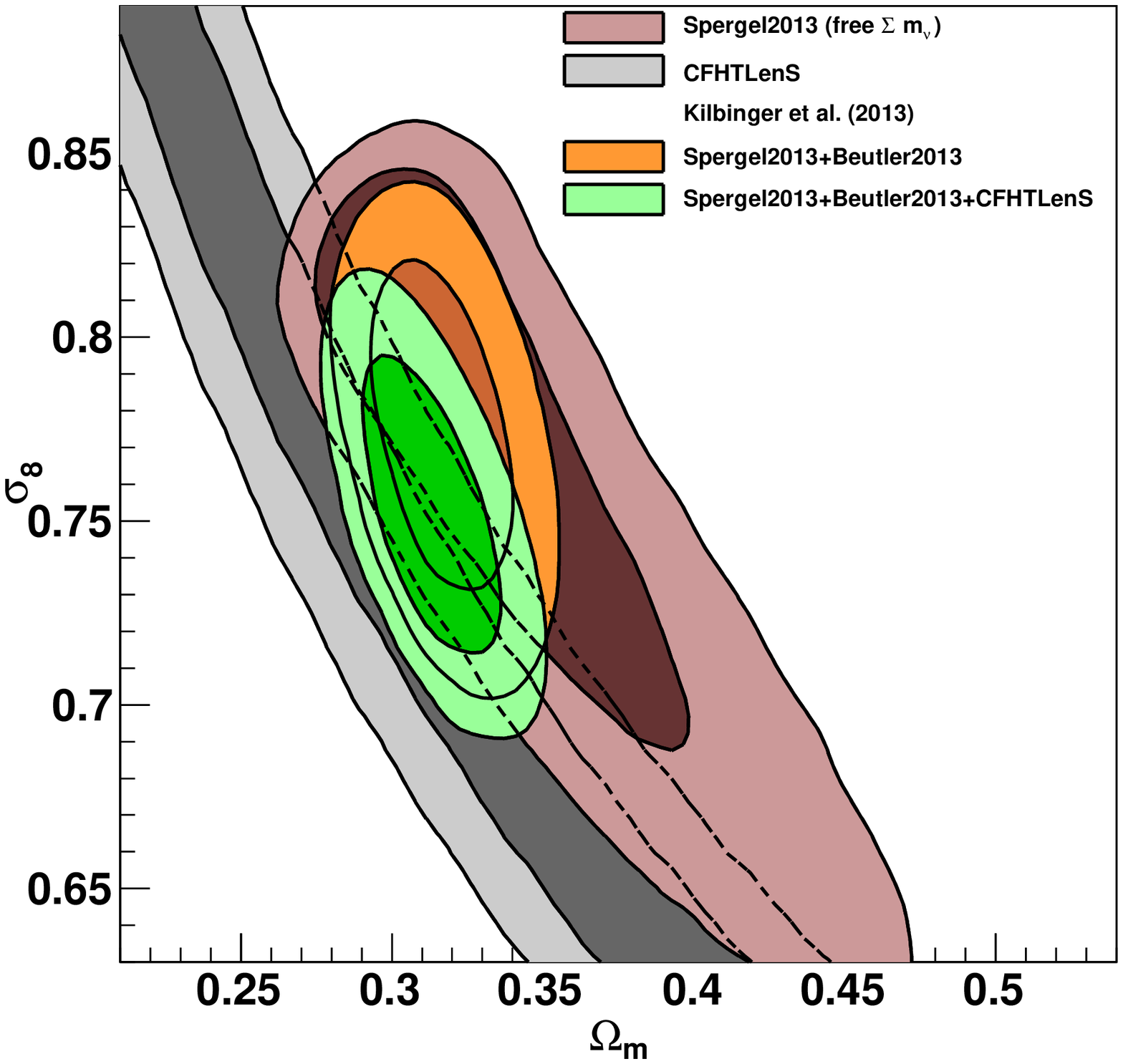,width=8.8cm}
\epsfig{file=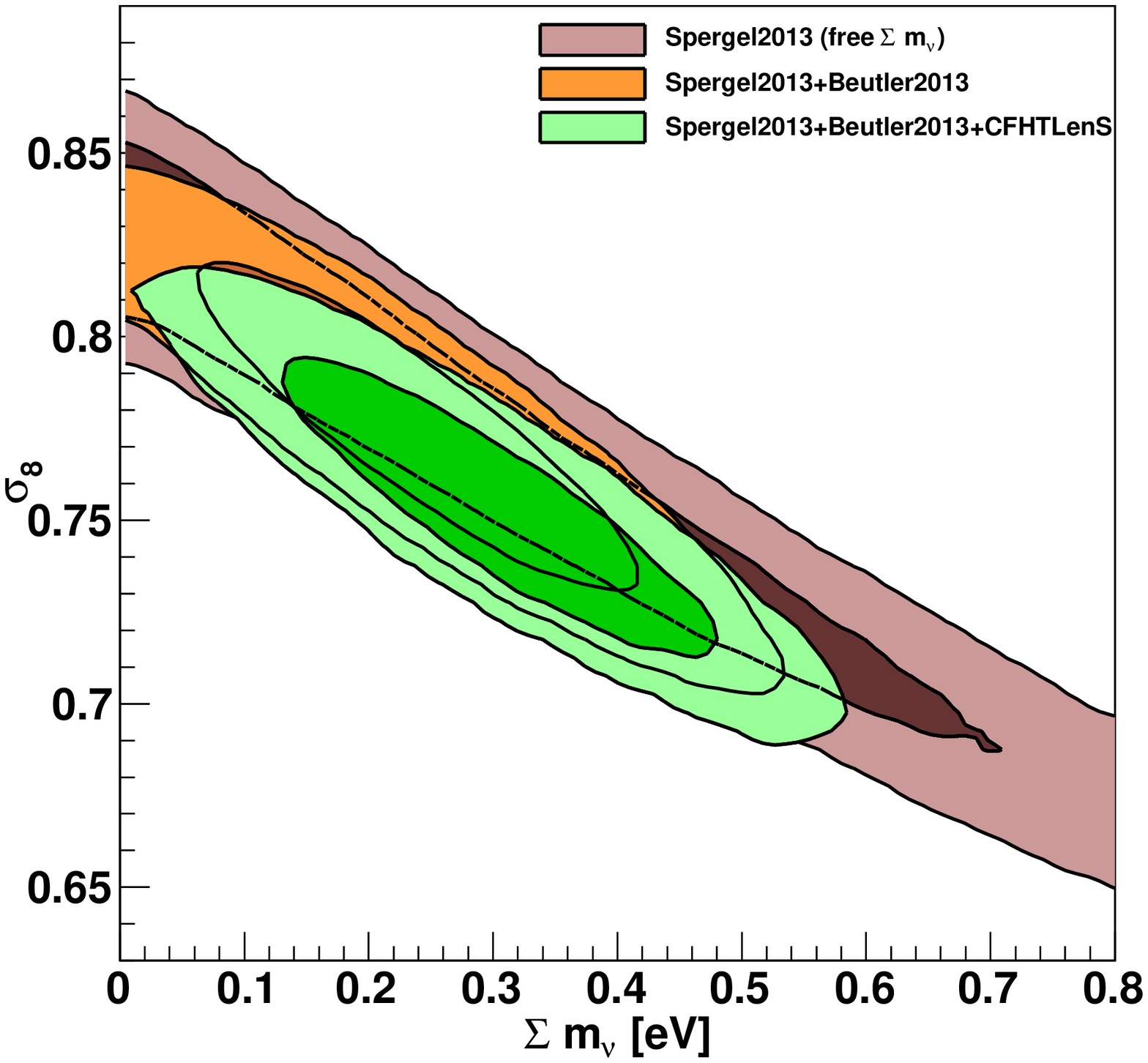,width=8.8cm}
\caption{Two-dimensional likelihood for $\Omega_m$-$\sigma_8$ (left) and $\sum m_{\nu}$-$\sigma_8$ (right) when combining the Planck re-analysis of~\citet{Spergel:2013rxa} within $\Lambda$CDM and free $\sum m_{\nu}$ with different low redshift growth of structure constraints. The orange contours show the Planck re-analysis combined with the $D_V/r_s$, $F_{\rm AP}$ and $f\sigma_8$ constraints of~\citet{Beutler:2013b}. The green contours additionally include CFHTLenS. The results are summarised in Table~\ref{tab:para2}.}
\label{fig:mnu3}
\end{center}
\end{figure*}

We now use the MCMC chains of~\citet{Spergel:2013rxa}, where the neutrino mass is varied freely and importance sample these chains. The chains we use include the A$_{\rm L}$-lensing signal, meaning they do not marginalise over $A_{\rm L}$. The CMB lensing signal from the $4$-point function is not included. The result is shown in Figure~\ref{fig:1Dprop2} (bottom), Figure~\ref{fig:mnu3} and Table~\ref{tab:para2}. Combining Spergel2013 with the results of~\citet{Beutler:2013b} yields $\sum m_{\nu} = 0.24\pm0.12\,$eV. 
Including CFHTLenS, GGlensing and further BAO constraints gives $\sum m_{\nu} = 0.29\pm0.10\,$eV ($2.9\sigma$). These results are within $1\sigma$ with the results we obtained when importance sampling the Planck and the Planck-$A_{\rm L}$ chains. Overall we see small (below $1\sigma$) shifts towards WMAP.

\section{Discussion}
\label{sec:dis}

We can summarise the results of the last section as follows:
\begin{enumerate}
\item We have a significant ($> 3\sigma$) detection of the neutrino mass when combining WMAP9 and Planck$-A_{\rm L}$ with low redshift growth of structure constraints. Planck$-A_{\rm L}$ represents the Planck dataset without the lensing contribution to the temperature power spectrum.
\item The $A_{\rm L}$-lensing contribution leads to $\sim 1\sigma$ shifts in $\Omega_m$ and $\sigma_8$, which have a non-negligible impact on the neutrino mass constraints.
\item When using Planck including the lensing contribution to the temperature power spectrum the significance of the detection of the neutrino mass is reduced to $\sim 2\sigma$.
\item The Planck re-analysis by~\citet{Spergel:2013rxa} shows results very similar to Planck with a slightly increased significance for the neutrino mass.
\end{enumerate}
It might not be too surprising that excluding the $A_{\rm L}$-lensing result brings Planck closer to WMAP9, since the $A_{\rm L}$-lensing contribution is unique to Planck and removing it increases the fraction of information common to the two datasets. Nevertheless, since there is a $2\sigma$ tension between the $A_{\rm L}$-lensing and the CMB lensing in the $4$-point function, it is interesting to examine the Planck data excluding the $A_{\rm L}$-lensing signal, especially given the shifts in $\sigma_8$ and $\Omega_m$, which significantly alter the constraints on $\sum m_{\nu}$.

The left panel in Figure~\ref{fig:mnu}, Figure~\ref{fig:mnu2} and Figure~\ref{fig:mnu3} show how the two dimensional constraints on $\Omega_m$ and $\sigma_8$ migrate when different datasets are included. The external datasets pull the combined constraint out of the $68\%$ confidence region of Planck (top panel of Figure~\ref{fig:mnu2}), indicating that increasing $\sum m_{\nu}$ does not resolve the tension between Planck and low redshift growth of structure constraints. WMAP9 and Planck$-A_{\rm L}$ present a different situation. Here the final constraints using all datasets lie within the $68\%$ confidence region of the CMB datasets. 

Figure~\ref{fig:1Dprop} and Figure~\ref{fig:1Dprop2} display the one-dimensional likelihood of $\sum m_{\nu}$ when combining low redshift growth of structure datasets with WMAP9, Planck, Planck$-A_{\rm L}$ and Spergel2013. While there is a prominent detection in the case of WMAP9 and Planck$-A_{\rm L}$, the detection in Planck is of low significance. The Planck re-analysis of~\citet{Spergel:2013rxa} shows a likelihood distribution very similar to the one obtained with the main Planck analysis.

We also note that there is tension between different components of the Planck dataset. While the amplitude of the $A_{\rm L}$-lensing in the $2$-point function prefers a small neutrino mass, the shape of the CMB lensing in the $4$-point function prefers a large neutrino mass. In Table~\ref{tab:para2} we can see that after marginalising over $A_{\rm L}$ the Planck dataset combined with CMB lensing prefers a neutrino mass of $0.86\,$eV with more than $2\sigma$.

To quantify the differences between the Planck and Planck$-A_{\rm L}$ chains used in this analysis we can look at $\Delta \chi^2$ for the best fitting cosmological parameters when combining the CMB datasets with~\citet{Beutler:2013b}, CFHTLenS, galaxy-galaxy lensing and the BAO constraints of 6dFGS and LOWZ. Within $\Lambda$CDM we find $\chi^2_{\rm Planck} - \chi^2_{\text{Planck}-A_{\rm L}} = 9.5$, while for $\Lambda$CDM$+\sum m_{\nu}$ we have $\Delta \chi^2 = 14.5$. In both cases the $\chi^2$ is reduced when excluding the $A_{\rm L}$-lensing contribution. We can also quantify which datasets drive the preference for neutrino mass. Considering the following dataset combinations (Planck, Planck+BAO, CFHTLenS+GGlensing, Beutler2013, Beutler2013$_{\rm BAO\-only}$), where BAO stands for the BAO constraints of 6dFGS and LOWZ, Beutler2013 stands for the three CMASS constraints ($D_V/r_s$, $F_{\rm AP}$, $f\sigma_8$) of~\citet{Beutler:2013b} and Beutler2013$_{\rm BAO\-only}$ represents only the BAO constraint ($D_V/r_s$). Note that $F_{\rm AP}$ and $f\sigma_8$ of~\citet{Beutler:2013b} are correlated and we can not easily explore the effect of just one of these constraints. We find the following $\Delta \chi^2$ values for the best fitting cosmology, when comparing $\Lambda$CDM and $\Lambda$CDM$+\sum m_{\nu}$: ($-4.3$, $-4.0$, $2.8$, $4.1$, $0.3$). The preference for neutrino mass in the case of Planck is driven by the CMASS and lensing constraints. If we instead use the best fitting cosmological parameters in case of Planck$-A_{\rm L}$ and again consider the dataset combinations (Planck$-A_{\rm L}$, Planck$-A_{\rm L}$+BAO, CFHTLenS+GGlensing, Beutler2013, Beutler2013$_{\rm BAO\-only}$) we find $\Delta \chi^2 = $ ($0.7$, $-0.17$, $3.9$, $4.8$, $0.5$). Again the preference for neutrino mass is driven by the CMASS and lensing constraints. Comparing Beutler2013 and Beutler2013$_{\rm BAO\-only}$ we see that it is mainly the RSD (and AP) constraints which drive the preference for neutrino mass within CMASS.

While in this paper we focus on the neutrino mass as a possible extension to $\Lambda$CDM, it is interesting to ask whether other parameters would also be able to alleviate the tension between the different datasets discussed in this paper. 
Comparing the best fitting $\chi^2$ of a universe with curvature as a free parameter (oCDM) to the best fitting $\chi^2$ when varying the neutrino mass we find $\Delta \chi^2 = \chi^2_{\rm oCDM} - \chi^2_{\sum m_{\nu}\Lambda CDM} = 6.27$, meaning that the neutrino mass is preferred as an extension to $\Lambda$CDM. Including instead a dark energy equation of state parameter (wCDM) we find $\Delta \chi^2 = \chi^2_{\rm wCDM} - \chi^2_{\sum m_{\nu}\Lambda CDM} = 0.68$, showing only mild preference for the neutrino mass parameter. If we include the neutrino mass as well as the number of relativistic species, $N_{\nu}$, as free parameters we find $\Delta \chi^2 = \chi^2_{\rm N_{\nu}\sum m_{\nu}\Lambda CDM} - \chi^2_{\sum m_{\nu}\Lambda CDM} = -0.99$. While the $\chi^2$ is reduced, the reduction of $\chi^2$ is not sufficient to justify the new parameter. We also note that the new parameter $N_{\nu}$ does not remove the preference for a non-zero neutrino mass, with the best fitting values being $N_{\nu} = 3.61\pm 0.35$ and $\sum m_{\nu} = 0.46\pm 0.18$.

Combining CMB datasets with external information on the Hubble parameter allows one to break the geometric degeneracy between $H_0$ and the neutrino mass parameter~\citep{Komatsu:2008hk} similar to the BAO constraints. A large neutrino mass leads to a smaller Hubble Constant and vice versa. Since the low redshift $H_0$ constraints using the distance ladder technique seem to find large values of $H_0$~\citep{Riess:2011yx,Freedman:2012ny,Efstathiou:2013via}  compared to CMB or BAO measurements, including the low redshift $H_0$ constraints usually does not lead to a detection of the neutrino mass (see e.g.~\citealt{Hou:2012xq,Verde:2013wza,Riemer-Sorensen:2013jsa,dePutter:2014hza,Zheng:2014dka}). 

\citet{Sanchez:2013tga} reported an upper limit on the neutrino mass of $\sum m_{\nu} < 0.23\,$eV ($95\%$ c.l.) using data from Planck, ACT and SPT, combined with CMASS-DR11. Their use of the CMASS-DR11 dataset is different to our analysis, since they make use of the shape of the correlation function wedges. Nevertheless, their result is consistent with our $95\%$ c.l. upper limit in Planck. When using WMAP9 instead of Planck,~\citet{Sanchez:2013tga} find $\sum m_{\nu}  = 0.23\pm 0.12\,$eV, which again is in $1\sigma$ agreement with our result.

A tight constraint on the neutrino mass of $\sum m_{\nu} < 0.17\,$eV ($95\%$ c.l.) has been reported in~\citet{Seljak:2006bg} by combining Ly-$\alpha$ forest power spectrum information with WMAP3 as well as supernovae and galaxy clustering constraints. One reason they achieved such a tight constraint was that their Ly-$\alpha$ forest measurement was in tension with WMAP3. Since this Ly-$\alpha$ forest measurement is now in good agreement with Planck, their upper limit would weaken. 

In addition to these non-detections, there are many studies which report a detection of the neutrino mass (see e.g.,~\citealt{Hou:2012xq,Ade:2013lmv,Wyman:2013lza,Rozo:2013hha}). \citet{Battye:2013xqa} showed that the Planck+CFHTLenS constraints are compatible with the constraints obtained when combining Planck with the SZ-cluster results also reported by the Planck team. They then performed an analysis combining Planck with BAO, CFHTLenS and SZ-clusters finding $\sum m_{\nu} = 0.320\pm0.081\,$eV, which is in good agreement with our results. This constraint is however dominated by the SZ-cluster constraint which suffers from various systematic errors~\citep{Costanzi:2013bha,Rozo:2012wy,Rozo:2013hha,vonderLinden:2014haa,Paranjape:2014lga}.

\subsection{Implications for General Relativity}
\label{sec:GR}

\begin{table*}
\begin{center}
\caption{Constraints on the growth index $\gamma$ and the sum of the neutrino masses from WMAP9 and Planck$-A_{\rm L}$ combined with the constraints of~\citet{Beutler:2013b}. For $\gamma$ we show $1\sigma$ errors, while for the sum of the neutrino masses we report the $68\%$ and $95\%$ confidence levels. The constraints on $\sum m_{\nu}$ are significantly degraded compared to the results in Table~\ref{tab:para2} because of the degeneracy with $\gamma$. The last two rows shows the constrains obtained in~\citet{Beutler:2013b} within $\Lambda$CDM for comparison.}
	\begin{tabular}{lccc}
		\hline
		dataset(s) & $\gamma$ & \multicolumn{2}{c}{$\sum m_{\nu}\;$[eV]}\\
		& & $68\%$ c.l. & $95\%$ c.l.\\
		\hline
		WMAP9+Beutler2013 & $0.72\pm 0.19$ & $0.47^{+0.23}_{-0.32}$ & $<0.85$ \\
		Planck$-A_{\rm L}$+Beutler2013 & $0.67\pm0.14$ & $0.25^{+0.13}_{-0.22}$ & $<0.52$ \\
		\hline
		WMAP9+Beutler2013 ($\Lambda$CDM) & $0.76\pm0.11$ & --- & ---\\
		Planck+Beutler2013 ($\Lambda$CDM) & $0.772^{+0.124}_{-0.097}$ & --- & --- \\
		\hline
	  \end{tabular}
	  \label{tab:para3}
\end{center}
\end{table*}

\begin{figure}
\begin{center}
\epsfig{file=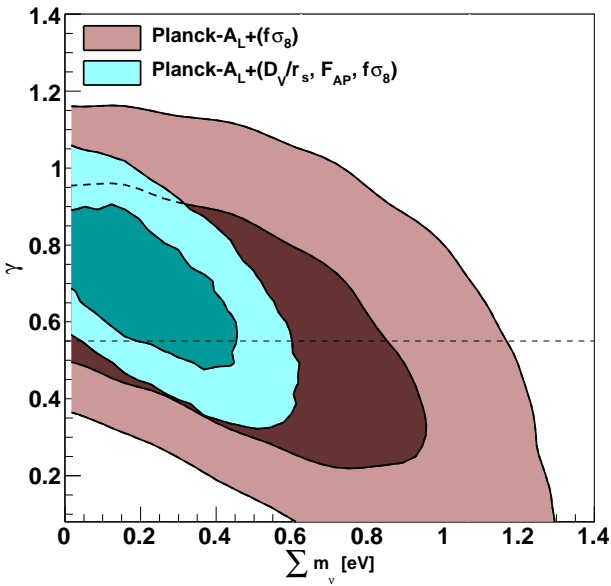,width=8.8cm}
\caption{Two-dimensional likelihood of the growth index $\gamma$ and $\sum m_{\nu}$. We combine different parts of the CMASS results of~\citet{Beutler:2013b} with Planck$-A_{\rm L}$. Planck$-A_{\rm L}$ is the Planck dataset where the $A_{\rm L}$-lensing contribution has been marginalised out. Combining Planck$-A_{\rm L}$ with the $D_V/r_s$, $F_{\rm AP}$ and $f\sigma_8$ constraint of~\citet{Beutler:2013b} (cyan contours) produces constraints on $\gamma$ in good agreement with the prediction by General Relativity (black dashed line).}
\label{fig:gamma_om}
\end{center}
\end{figure}

\begin{figure}
\begin{center}
\epsfig{file=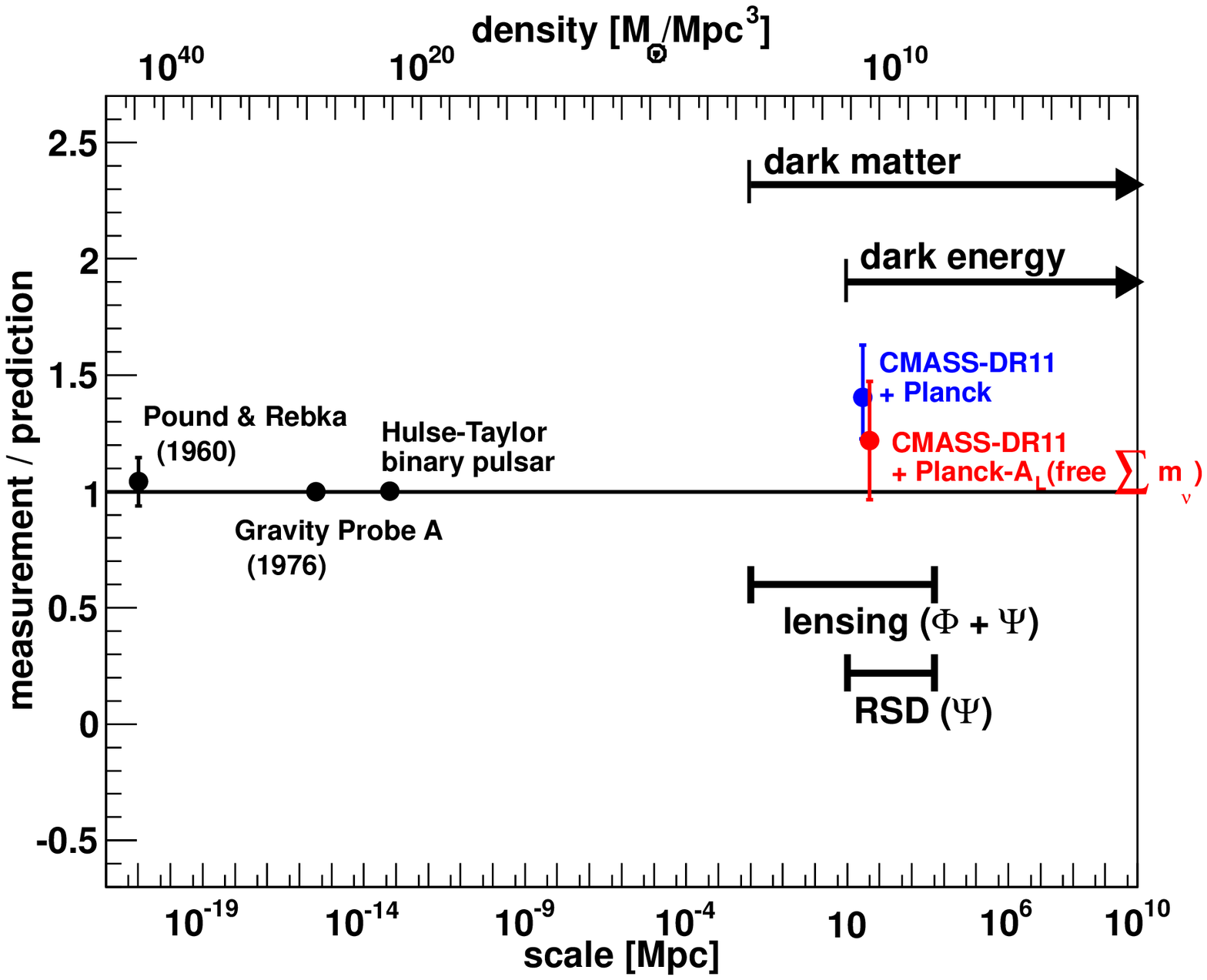,width=8.8cm}
\caption{Summary of different tests of General Relativity (GR) as a function of distance scale (bottom axis) and densities (top axis). The figure includes the Pound \& Rebka experiment~\citep{Pound:1960zz}, Gravity Probe A~\citep{Vessot:1980zz} and the Hulse-Taylor binary pulsar~\citep{Hulse:1974eb}. The error bars for Gravity Probe A and the Hulse-Taylor binary pulsar are smaller than the data points in this plot. In blue we include the result of~\citet{Beutler:2013b}, where Planck (within $\Lambda$CDM) has been combined with CMASS-DR11 constraints, finding a $2\sigma$ tension. In this analysis we use the Planck result where the $A_{\rm L}$-lensing contribution has been marginalised out and vary $\sum m_{\nu}$ (red datapoint).}
\label{fig:keff}
\end{center}
\end{figure}

\citet{Beutler:2013b} combined CMASS-DR11 results with Planck and WMAP9 to test General Relativity (GR) using the simple $\gamma$-parameterisation, where the growth rate is given by $f(z) \simeq \Omega^{\gamma}_m(z)$~\citep{Linder:2005in}. \citet{Beutler:2013b} found $\gamma = 0.772^{+0.124}_{-0.097}$ when combining with Planck and $\gamma = 0.76\pm0.11$ when combining with WMAP9. These results are in $2\sigma$ tension with the GR prediction of $\gamma^{\rm GR} \approx 0.55$. The question is now, what are the implications of a neutrino mass for these results? 

As discussed in~\citet{Beutler:2013b}, the tension with GR is mainly caused by the large $\sigma_8$ in the CMB datasets. Since introducing a neutrino mass reduces the CMB prediction of $\sigma_8$, we can expect that a non-zero neutrino mass will also decrease the tension with GR. The reason to combine the clustering result of CMASS with a CMB dataset is the need to add information on $\sigma_8$ to be able to test gravity through the growth rate $f(z)$. Since the uncertainty in $\sigma_8$ significantly increases when the neutrino mass is varied freely, we expect that the error on $\gamma$ will increase as well.

Here we use the two CMB chains with the strongest signs of a neutrino mass, which are WMAP9 and Planck$-A_{\rm L}$. We use the chains which have the sum of the neutrino masses as a free parameter. We importance sample these chains and include $\gamma$ as an additional free parameter following the procedure of section 9.1 in~\citet{Beutler:2013b}. Marginalising over all other parameters we find $\gamma = 0.72\pm0.19$ for WMAP9 and $\gamma = 0.67\pm0.14$ for Planck$-A_{\rm L}$. Both results are in $1\sigma$ agreement with the GR prediction. The result for the Planck$-A_{\rm L}$ chain is shown in Figure~\ref{fig:gamma_om} and Figure~\ref{fig:keff}. Even though the constraints on the sum of the neutrino masses for this test are significantly degraded, because of the degeneracy with $\gamma$, we include them in Table~\ref{tab:para3}. Figure~\ref{fig:keff} compares the result of this analysis (red datapoint) with the result in~\citet{Beutler:2013b} (blue datapoint). It might not be surprising that the tension with GR in~\citet{Beutler:2013b} can be reduced by introducing a new parameter, especially if this parameter is degenerate with $\gamma$. 

\subsection{Implications for particle physics}
\label{sec:particle}

Although our evidence of the neutrino mass has to be taken with care given the significance of the detection ($\sim 2.5$- $3.5\sigma$) and the tension with the $A_{\rm L}$-lensing contribution to the Planck measurement, it is still interesting to investigate the implications of such a result.

What are the implications for the masses of the neutrino eigenstates? We use the mass difference $|\Delta m^2_{31}| = 2.4\times 10^{-3}$eV$^2$~\citep{Beringer:1900zz} and our measurement $\sum m_{\nu} = 0.36\pm0.10\,$eV, which was obtained by combining Planck$-A_{\rm L}$ with~\citet{Beutler:2013b}, CFHTLenS, galaxy-galaxy lensing and BAO constraints from 6dFGS and LOWZ. If we further assume three neutrinos arranged by the normal hierarchy with the two light neutrinos ($m_{\nu_{1,2}}$) having the same mass, we find $m_{\nu_3} = 0.127\pm0.032\,$eV and $m_{\nu_{1,2}} = 0.117\pm 0.032\,$eV. For the inverted hierarchy we get instead $m_{\nu_{1,2}} = 0.123\pm 0.032\,$eV and $m_{\nu_3} = 0.113\pm0.032\,$eV. 

Given a certain hierarchy we can calculate the flavour eigenstates using the mixing matrix (Pontecorvo-Maki-Nakagawa-Sakata matrix)
\begin{equation}
U_{\rm PMNS} = 
\left(\begin{matrix}0.82 & 0.55 & 0.15\cr
-0.50 & 0.58 & 0.64\cr
0.26 & -0.60 & 0.75\end{matrix}\right),
\end{equation}
where we assume any possible complex phase to be zero and use the mixing angles from~\citet{Beringer:1900zz} and~\citet{An:2013zwz}. The flavour eigenstates are then given as super-position of the mass eigenstates
\begin{equation}
\left(\begin{matrix}|\nu_e\rangle\cr
|\nu_{\mu}\rangle\cr
|\nu_{\tau}\rangle\end{matrix}\right) = U_{\rm PMNS}\times \left(\begin{matrix}|\nu_1\rangle\cr
|\nu_2\rangle\cr
|\nu_3\rangle\end{matrix}\right).
\end{equation}
Because of neutrino mixing, the observable of different direct neutrino mass experiments is different to the flavour states. Neutrino-less double beta decay ($0\nu\beta\beta$) experiments are sensitive to the mass
\begin{equation}
m_{\beta\beta} = \sum^3_{i=1} m_{\nu_i} U^2_{{\rm PMNS}, 1i},
\end{equation}
while beta-decay experiments are sensitive to 
\begin{equation}
m_{\beta} = \sqrt{\sum^3_{i=1} m^2_{\nu_i} U^2_{{\rm PMNS}, 1i}}.
\label{eq:mbeta}
\end{equation}
Taking the constraints on the mass eigenstates above together with the mixing matrix we find $m_{\beta} = 0.117\pm0.031\,$eV for the normal hierarchy and $m_{\beta} = 0.123\pm0.032\,$eV for the inverted hierarchy\footnote{The errors on the mixing angles are not propagated, since the error budget is dominated by the uncertainty in the sum of the neutrino masses.}. Since the masses are close to degenerate and because $U_{{\rm PMNS}, 13}$ is small compared to $U_{{\rm PMNS}, 11}$ and $U_{{\rm PMNS}, 12}$, the values of $m_{\beta}$ are basically identical to $m_{\nu_{1,2}}$. The value of $m_{\beta}$ in both hierarchies is below the predicted sensitivity range of the KATRIN experiment.

\section{Conclusion}
\label{sec:conclusion}

This paper presents an investigation of the cosmological implications of the CMASS-DR11 anisotropic analysis including the growth of structure measurement, with particular focus on the sum of the neutrino masses $\sum m_{\nu}$. 

First we examine the robustness of the CMASS constraints of~\citet{Beutler:2013b} when changing the power spectrum template including $\sum m_{\nu} = 0.4\,$eV. Our main cosmological parameters change by $<0.5\sigma$ and therefore are robust against variations in the neutrino mass. We perform similar tests for the weak lensing results from CFHTLenS, finding that these results show only weak dependence on the initial assumption of the neutrino mass parameter.

We use the WMAP9 and Planck MCMC chains where the sum of the neutrino masses is varied as a free parameter and importance sample these chains. When combining WMAP9 with the three constraints $(D_V/r_s,F_{\rm AP},f\sigma_8)$ of~\citet{Beutler:2013b} we obtain $\sum m_{\nu} = 0.36\pm0.14\,$eV, which represents a $2.6\sigma$ preference for the neutrino mass. If we also include CFHTLenS, galaxy-galaxy lensing and the BAO constraints from 6dFGS and LOWZ, we find $\sum m_{\nu} = 0.35\pm0.10\,$eV ($3.3\sigma$). 

Using the Planck dataset the preference for a neutrino mass is reduced to $\sim 2\sigma$. However, marginalising over the $A_{\rm L}$-lensing contribution to the temperature power spectrum of Planck leads to $\sim 1\sigma$ shifts in $\Omega_m$ and $\sigma_8$, which bring Planck into much better agreement with WMAP9. Combining Planck without the $A_{\rm L}$-lensing contribution with CMASS yields similar results to WMAP9. We find $\sum m_{\nu} = 0.36\pm0.10\,$eV ($3.4\sigma$) when combining with~\citet{Beutler:2013b}, CFHTLenS, galaxy-galaxy lensing and further BAO constraints. This constraint is robust against various permutations of datasets (see Table~\ref{tab:para2} for details). We also investigated the Planck re-analysis of~\citet{Spergel:2013rxa}, finding that it yields results very similar to Planck with a slightly increased significance for a neutrino mass. While the preference for neutrino mass is driven mainly by the low redshift growth of structure constraints it is reassuring that the three growth of structure datasets included in this analysis (CMASS-RSD, CFHTLenS and galaxy-galaxy lensing) yield consistent results. Our constraints could be significantly improved by including cluster counts detected through the Sunyaev-Zeldovich effect. We chose, however, to not include these datasets, because of the significant systematic uncertainty of these measurements with respect to the treatment of the neutrino mass. 

In this paper we present many combinations of datasets and a natural question is, which of these presents the main result of this paper. When discussing the implications of our results in section~\ref{sec:particle}, we selected the constraint $\sum m_{\nu} = 0.35\pm0.10\,$eV, obtained with the Planck$-A_{\rm L}$ chain. However, we cannot conclusively put this forward as the fiducial result of our analysis, without having an explanation for the tension with the $A_{\rm L}$-lensing amplitude. The origin of the tension between the different components in the Planck dataset remains an open question, which we will hopefully learn more about with the next data release of Planck.

A neutrino mass at this level would relieve the tension of current datasets with the clustering prediction of GR reported in~\citet{Beutler:2013b}. If we remove the $A_{\rm L}$-lensing contribution to Planck and combine with the $(D_V/r_s,F_{\rm AP},f\sigma_8)$ constraints of~\citet{Beutler:2013b} by varying the neutrino mass and the growth index $\gamma$ as free parameters, where $f(z) = \Omega_m^{\gamma}(z)$, we find $\gamma = 0.67\pm0.14$. This result is in $1\sigma$ agreement with the GR prediction of $\gamma^{\rm GR} = 0.55$. Similar results are obtained for WMAP9.

If our result is confirmed by future, more precise cosmological measurements, it will have significant implications for particle physics and cosmology. The constraint $\sum m_{\nu} = 0.35\pm0.10\,$eV can be expressed as
\begin{align}
\Omega_{\nu}h^2 &= 0.0039\pm0.0011\;\;\;\;\; \text{or}\\
f_{\nu} &= 0.0315\pm0.0088.
\end{align}
The large value of $\sum m_{\nu}$ found in our analysis would be too large to allow for cosmological probes to distinguish between the inverted and the normal mass hierarchies just by the measurement of the sum of the masses. However, within the normal hierarchy we can predict $m_{\beta} = 0.117\pm0.031\,$eV, while for the inverted hierarchy we find $m_{\beta} = 0.123\pm0.032\,$eV. These masses are below the predicted detection limits of the KATRIN experiment (assuming a sensitivity of $m_{\beta} \sim 0.2\,$eV~\citealt{Wolf:2008hf}). 

The constraints presented in this paper will be further improved in the near future. 
Within the CMASS dataset the weakest point of the constraint is certainly the large uncertainty on $f\sigma_8$, which, however, is predicted to improve significantly with future datasets like the Dark Energy Spectroscopic Instrument (DESI)~\citep{Font-Ribera:2013rwa, Abazajian:2013oma}. Even the BOSS dataset could provide additional constraints on the neutrino mass using the characteristic scale-dependent damping of the power spectrum (see~\citealt{Zhao:2012xw} for such an attempt) which, however, requires refined simulations including massive neutrinos~\citep{Villaescusa-Navarro:2013pva}. 

\section*{Acknowledgments}

We would like to thank Renee Hlozek for providing the MCMC chains for the Planck re-analysis. FB would like to thank Martin Kilbinger and Catherine Heymans for help with the CFHTLenS dataset and CosmoPMC. FB would also like to thank Martin White, Uros Seljak, Eric Linder, Daniel Dwyer, Morag Scrimgeour, Michael Mortonson, Marcel Schmittful and Blake Sherwin for helpful discussion. SS would like to thank Kiyotomo Ichiki and Masahiro Takada for providing their MCMC code for weak lensing analysis and for useful discussions. SS is supported by a Grant-in-Aid for Young Scientists (Start-up) from the Japan Society for the Promotion of Science (JSPS) (No. 25887012).

Funding for SDSS-III has been provided by the Alfred P. Sloan Foundation, the Participating Institutions, the National Science Foundation, and the U.S. Department of Energy Office of Science. The SDSS-III web site is http://www.sdss3.org/.

SDSS-III is managed by the Astrophysical Research Consortium for the Participating Institutions of the SDSS-III Collaboration including the University of Arizona, the Brazilian Participation Group, Brookhaven National Laboratory, Carnegie Mellon University, University of Florida, the French Participation Group, the German Participation Group, Harvard University, the Instituto de Astrofisica de Canarias, the Michigan State/Notre Dame/JINA Participation Group, Johns Hopkins University, Lawrence Berkeley National Laboratory, Max Planck Institute for Astrophysics, Max Planck Institute for Extraterrestrial Physics, New Mexico State University, New York University, Ohio State University, Pennsylvania State University, University of Portsmouth, Princeton University, the Spanish Participation Group, University of Tokyo, University of Utah, Vanderbilt University, University of Virginia, University of Washington, and Yale University.

This research used resources of the National Energy Research Scientific Computing Center, which is supported by the Office of Science of the U.S. Department of Energy under Contract No. DE-AC02-05CH11231.

\setlength{\bibhang}{2em}
\setlength{\labelwidth}{0pt}

\newpage

\appendix
\numberwithin{equation}{section}

\newpage

\label{lastpage}

\end{document}